\documentclass{article}
\usepackage{amsmath}
\usepackage{graphics}
\usepackage{color}

\begin{document}

\title{Conceptual tensions between quantum mechanics and general relativity:
Are there experimental consequences, e.g., superconducting transducers
between electromagnetic and gravitational radiation?}
\author{Raymond Y. Chiao \\
%EndAName
Department of Physics\\
University of California\\
Berkeley, CA 94720-7300\\
U. S. A.\\
(E-mail: chiao@physics.berkeley.edu)}
\date{Chapter for the Wheeler Volume of September 17, 2002 }
\maketitle

\begin{abstract}
One\ of the conceptual tensions between quantum mechanics (QM) and general
relativity (GR) arises from the clash between the \textit{spatial
nonseparability} of entangled states in QM, and the complete \textit{spatial
separability} of all physical systems in GR, i.e., between the $nonlocality$
implied by the superposition principle, and the $locality$ implied by the
equivalence principle. \ Experimental consequences of this conceptual
tension will be explored for macroscopically coherent quantum fluids, such
as superconductors, superfluids, and atomic Bose-Einstein condensates
(BECs), subjected to tidal and Lense-Thirring fields arising from
gravitational radiation. \ A Meissner-like effect is predicted, in which the
Lense-Thirring field is expelled from the bulk of a quantum fluid, apart
from a thin layer given by the London penetration depth $\lambda _{L}$. \
For atomic BECs, $\lambda _{L}$ is a microscopic length given by $(8\pi 
\overline{n}a)^{-1/2}$ where $\overline{n}$ is the mean atomic density of
the BEC, and $a$ is the scattering length of $S$-wave scattering, which
produces quantum entanglement of pairs of atoms with opposite momenta. \
Superconductors are predicted to be macroscopic quantum gravitational
antennas and transducers, which can directly convert upon reflection a beam
of quadrupolar electromagnetic radiation into gravitational radiation, and
vice versa, and thus serve as both sources and receivers of gravitational
waves. \ An estimate of the transducer conversion efficiency on the order of
unity comes out of the Ginzburg-Landau theory for an extreme type II,
dissipationless superconductor with minimal coupling to weak gravitational
and electromagnetic radiation fields, whose frequency is smaller than the
BCS gap frequency, thus satisfying the quantum adiabatic theorem. \ The
concept of ``the impedance of free space for gravitational plane waves'' is
introduced, and leads to a natural impedance-matching process, in which the
two kinds of radiation fields are impedance-matched to each other around a
hundred coherence lengths beneath the surface of the superconductor. \ A
simple, Hertz-like experiment has been performed to test these ideas, and
preliminary results will be reported.
\end{abstract}

\section{Introduction}

\begin{quotation}
\bigskip ``Mercy and Truth are met together; Righteousness and Peace have
kissed each other.'' \ (Psalm 85:10)
\end{quotation}

In this Festschrift Volume in honor of John Archibald Wheeler, I would like
to take a fresh look at the intersection between two fields to which he
devoted much of his research life: general relativity (GR) and quantum
mechanics (QM). As evidence of his keen interest in these two subjects, I
would cite two examples from my own experience. When I was an undergraduate
at Princeton University during the years from 1957 to 1961, he was my
adviser. One of his duties was to assign me topics for my junior paper and
for my senior thesis. For my junior paper, I was assigned the topic: Compare
the complementarity and the uncertainty principles of quantum mechanics:
Which is more fundamental? For my senior thesis, I was assigned the topic:
How to quantize general relativity? As Wheeler taught me, more than half of
science is devoted to the asking of the right question, while often less
than half is devoted to the obtaining of the correct answer, but not always!
\ 

In the same spirit, I would like to offer up here some questions concerning
conceptual tensions between GR and QM, which hopefully can be answered in
the course of time by experiments, and, in particular, to make some specific
experimental proposals, such as the use of macroscopic quantum objects like
superconductors, as quantum transducers and antennas for gravitational
radiation, in order to probe the tension between the concepts of $locality$
in GR and $nonlocality$ in QM.

\begin{figure}[tbp]
\centerline{\includegraphics{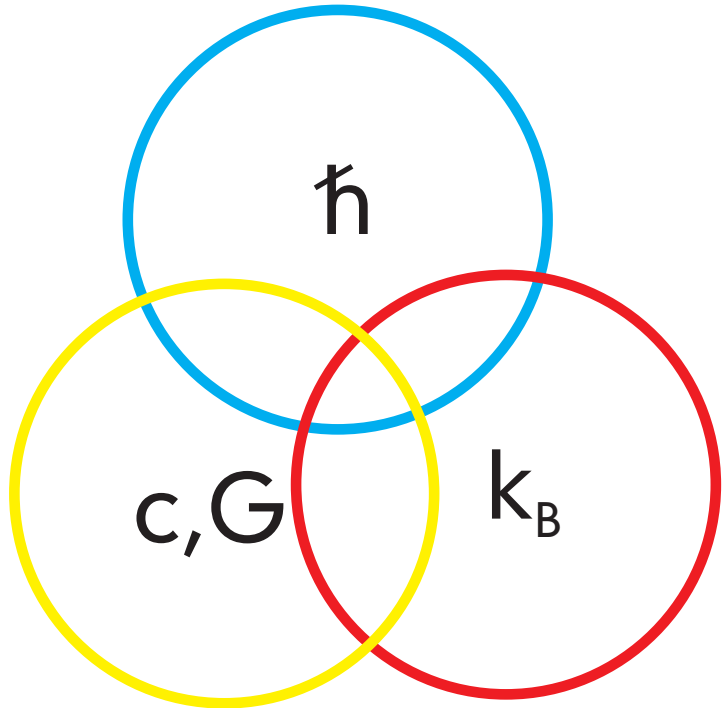}}
\caption{Three intersecting circles in a Venn-like diagram represent three
conceptual domains, which constitute the three main pillars of physics at
the beginning of the 21st century. \ The top circle represents quantum
mechanics, and is labeled by Planck's constant $\hbar $. \ The left circle
represents relativity, and is labeled by the two constants $c$, the speed of
light, and $G$, Newton's constant. \ The right circle represents statistical
mechanics and thermodynamics, and is labeled by Boltzmann's constant $k_{B}$%
. \ Conceptual tensions exist at the intersections of these three circles,
which may lead to fruitful experimental consequences.}
\label{threecircles}
\end{figure}

As I see it, the three main pillars of physics at the beginning of the 21st
century are quantum mechanics, relativity, and statistical mechanics, which
correspond to Einstein's three papers of 1905. There exist conceptual
tensions at the intersections of these three fields of physics (see Figure %
\ref{threecircles}). It seems worthwhile re-examining these tensions, since
they may lead to new experimental discoveries. In this introduction, I shall
only briefly mention three conceptual tensions between these three fields: $%
locality$ versus $nonlocality$ of physical systems, $objectivity$ versus $%
subjectivity$ of probabilities in quantum and statistical mechanics (the
problem of the nature of information), and $reversibility$ versus $%
irreversibility$ of time (the problem of the arrows of time). \ Others in
this Volume will discuss the second and\ the third of these tensions in
detail. \ I shall limit myself to a discussion of the first conceptual
tension concerning locality versus nonlocality, mainly in the context of GR
and QM. \ (However, in my Solvay lecture \cite{ChiaoSolvay}, I have
discussed the other two tensions in more detail).

Why examine conceptual tensions? \ A brief answer is that they often lead to
experimental discoveries. \ It suffices to give just one example from late
19th and early 20th century physics: the clash between the venerable
concepts of $continuity$ and $discreteness$. \ The concept of continuity,
which goes back to the Greek philosopher Heraclitus (``everything flows''),
clashed with the concept of discreteness, which goes back to Democritus
(``everything is composed of atoms''). \ Eventually, Heraclitus's concept of
continuity, or more specifically that of the $continuum$, was embodied in
the idea of $field$ in the classical field theory associated with Maxwell's
equations. \ The atomic hypothesis of Democritus was eventually embodied in
the kinetic theory of gases in statistical mechanics. \ 

Conceptual tensions, or what Wheeler calls the ``clash of ideas,'' need not
lead to a complete victory of one conflicting idea over the other, so as to
eliminate the opposing one completely, as seemed to be the case in the 19th
century, when Newton's idea of ``corpuscles of light'' was apparently
completely eliminated in favor of the wave theory of light. \ Rather, there
may result a reconciliation of the two conflicting ideas, which\ then often
leads to many fruitful experimental consequences. \ 

Experiments on blackbody radiation in the 19th century were exploring the
intersection, or borderline, between Maxwell's theory of electromagnetism
and statistical mechanics, where the conceptual tension between continuity
and discreteness was most acute, and eventually led to the discovery of
quantum mechanics through the work of Planck. The concept of $discreteness$
metamorphosed into the concept of the $quantum$. \ This led\ in turn to the
concept of \textit{discontinuity} embodied in Bohr's \textit{quantum jump}
hypothesis, which was necessitated by the indivisibility of the quantum.
Many experiments, such as Millikan's measurements of $h/e$, were in turn
motivated by Einstein's heuristic theory of the photoelectric effect based
on the quantum hypothesis. \ Newton's idea of ``corpuscles of light''
metamorphosed into the concept of the $photon$. \ This is a striking example
showing how that many fruitful experimental consequences can come out of one
particular conceptual tension.

Within a broader cultural context, there have been many acute conceptual
tensions between science and faith, which have lasted over many centuries. \
Perhaps the above examples of the fruitfulness of the resolution of
conceptual tensions within physics itself may serve as a parable concerning
the possibility of a peaceful reconciliation of these great cultural
tensions, which may eventually lead to the further growth of both science
and faith. \ Hence we should not shy away from conceptual tensions, but
rather explore them with an honest, bold, and open spirit.

\section{Three conceptual tensions between quantum mechanics and general
relativity}

Here I shall focus my attention specifically on some conceptual tensions at
the intersection between QM and GR. \ A commonly held viewpoint within the
physics community today is that the only place where conceptual tensions
between these two fields can arise is at the microscopic Planck length
scale, where quantum fluctuations of spacetime (``quantum foam'') occur. \
Hence manifestations of these tensions would be expected to occur only in
conjunction with extremely high-energy phenomena, accessible presumably only
in astrophysical settings, such as the Big Bang. \ \qquad

However, I believe that this point of view is too narrow. \ There exist
other\ conceptual tensions at macroscopic, non-Planckian\ distance scales,
which should be accessible in low-energy laboratory experiments involving
macroscopic QM phenomena. \ It should be kept in mind that QM not only
describes $microscopic$ phenomena, but also $macroscopic$ phenomena, such as
superconductivity. \ Specifically, I would like to point out the following\
three conceptual tensions:

\begin{description}
\item (1) The \textit{spatial nonseparability} of physical systems due to
entangled states in QM, versus the complete \textit{spatial separability} of
all physical systems in GR.

\item (2) The \textit{equivalence principle} of GR, versus the \textit{%
uncertainty principle} of QM.

\item (3) The \textit{mixed state} (e.g., of an entangled bipartite system,
one part of which falls into a black hole; the other of which flies off to
infinity) in GR, versus the \textit{pure state} of such a system in QM.
\end{description}

Conceptual tension (3) concerns the problem of the nature of information in
QM and GR. \ Again, since others will discuss this tension in detail in this
Volume, I shall limit myself only to a discussion of the first two of these
tensions. \ 

These conceptual tensions originate from the \textit{superposition principle}
of QM, which finds its most dramatic expression in the \textit{entangled
state} of two or more spatially separated particles of a single physical
system, which in turn leads to Einstein-Podolsky-Rosen\ (EPR) effects. \ It
should be emphasized here that it is necessary to consider \textit{two or
more} particles for observing EPR phenomena, since only then does the 
\textit{configuration space} of these particles no longer coincide with that
of ordinary spacetime. \ For example, consider the entangled state of two
spin 1/2 particles in a Bohm singlet state initially prepared in a total
spin zero state 
\begin{equation}
\left| S=0\right\rangle =\frac{1}{\sqrt{2}}\left\{ \left| \uparrow
\right\rangle _{1}\left| \downarrow \right\rangle _{2}-\left| \uparrow
\right\rangle _{2}\left| \downarrow \right\rangle _{1}\right\} ,
\label{Bohm singlet}
\end{equation}%
in which the two particles in a spontaneous decay process fly arbitrarily
far away from each other into two space-like separated regions of spacetime,
where measurements on spin by means of two Stern-Gerlach apparati are
performed separately on these two particles. \ 

As a result of the quantum entanglement arising from the $superposition$ of $%
product$ states, such as in the above Bohm singlet state, it is in general
impossible to factorize this state into products of probability amplitudes.
\ Hence it is impossible to factorize the $joint$ probabilities in the
measurements of spin of this two-particle system. \ This mathematical $%
nonfactorizability$ implies a physical $nonseparability$ of the system, and
leads to instantaneous, space-like correlations-at-a-distance in the joint
measurements of the properties (e.g., spin) of discrete events, such as in
the coincidence detection of ``clicks'' in Geiger counters placed behind the
two distant Stern-Gerlach apparati. \ These long-range correlations violate
Bell's inequalities, and therefore cannot be explained on the basis of any 
\textit{local realistic} theories. \ 

Violations of Bell's inequalities have been extensively experimentally
demonstrated \cite{Gisin2002}. \ These violations were predicted by QM. \ If
we assume a realistic world view, i.e., that the ``clicks'' of the Geiger
counters really happened, then we must conclude that we have observed $%
nonlocal$ features of the world. \ Therefore, a fundamental \textit{spatial
nonseparability} of physical systems has been revealed by these
Bell-inequalities-violating EPR correlations \cite{Chiao and Garrison}. \ It
should be emphasized that these space-like EPR correlations occur on
macroscopic, non-Planckian distance scales, where the conceptual tension (1)
between QM and GR becomes most acute.

Gravity is a \textit{long-range} force. \ It is therefore\ natural\ to
expect that experimental consequences of conceptual tension (1)\ should
manifest themselves most dramatically in the\ interaction of macroscopic
quantum matter, which exhibit \textit{long-range} EPR correlations, with 
\textit{long-range} gravitational fields. \ In particular, the question
naturally arises: How do entangled states, such as the Bohm singlet state,
interact with tidal fields, such as those in gravitational radiation \cite%
{Taylor}? \ It is therefore natural to look to the realm of\ \textit{%
macroscopic}, long-range quantum phenomena, such as superconductivity,
rather than phenomena at \textit{microscopic}, Planck length scales, in our
search for these experimental consequences.

Already a decade or so before Bell's ground-breaking work on his famous
inequality, Einstein himself was clearly worried by the radical, spatial
nonseparability of physical systems in quantum mechanics. \ Einstein wrote %
\cite{Einstein-Born}:

\begin{quotation}
``Let us consider a physical system S$_{12}$, which consists of two
part-systems S$_{1}$ and S$_{2}$. \ These two part-systems may have been in
a state of mutual physical interaction at an earlier time. \ We are,
however, considering them at a time when this interaction is at an end. \
Let the entire system be completely described in the quantum mechanical
sense by a $\mathrm{\psi }$-function $\mathrm{\psi }_{12}$ of the
coordinates $\mathrm{q}_{1}$,... and $\mathrm{q}_{2}$,... of the two
part-systems ($\mathrm{\psi }_{12}$ cannot be represented as a product of
the form $\mathrm{\psi }_{1}\mathrm{\psi }_{2}$ but only as a sum of such
products [\textit{i.e., as an entangled state}]). \ At time\textrm{\ }t%
\textrm{\ }let the two part-systems be separated from each other in space,
in such a way that $\mathrm{\psi }_{12}$ only differs from zero when $%
\mathrm{q}_{1}$,... belong to a limited part R$_{1}$ of space and \ $\mathrm{%
q}_{2}$,... belong to a part R$_{2}$ separated from R$_{1}$. \ \ \ . . .

``There seems to me no doubt that those physicists who regard the
descriptive methods of quantum mechanics as definitive in principle would
react to this line of thought in the following way: they would drop the
requirement for \textit{the independent existence of the physical reality
present in different parts of space}; they would be justified in pointing
out that the quantum theory nowhere makes explicit use of this
requirement.'' [\textit{Italics mine.}]
\end{quotation}

This radical, \textit{spatial nonseparability} of a physical system
consisting of two or more entangled particles in QM, which seems to
undermine the very possibility of the concept of \textit{field} in physics,
is in an obvious conceptual tension with the\ complete spatial separability
of any physical system into\ its separate parts in GR, which is a \textit{%
local realistic} field theory. \ 

However, I should hasten to add immediately that the battle-tested concept
of \textit{field} has of course been extremely fruitful not only at the
classical but also at the quantum level. \ Relativistic quantum field
theories have been very well validated, at least in an approximate,
correspondence-principle sense in which spacetime itself is treated
classically, i.e., as being describable by a rigidly flat, Minkowskian
metric, which has no\ possibility of any quantum dynamics. \ There have been
tremendous successes of quantum electrodynamics and electroweak gauge field
theory (and, to a lesser extent, quantum chromodynamics) in passing all
known high-energy experimental tests. \ Thus the conceptual tension between $%
continuity$ (used in the concept of the spacetime $continuum$) and $%
discreteness$ (used in the concept of \textit{quantized excitations} of a
field in classical spacetime) seems to have been successfully reconciled in
these relativistic quantum field theories. \ Nevertheless, the problem of a
satisfactory relativistic treatment of quantum measurement within these
theories remains an open one.

\subsection{Quantum fluids versus perfect fluids \ }

Here I shall propose some low-energy experimental probes of conceptual
tension (1), using \textit{macroscopically entangled}, and thus \textit{%
radically delocalized}, quantum states encountered in large quantum objects,
such as superconductors, superfluids, and the recently observed atomic
Bose-Einstein condensates (BECs), i.e., in what I shall henceforth call
``quantum fluids.'' \ Again it should be stressed that gravity is a \textit{%
long-range} force, and therefore it should be possible to perform \textit{%
low-energy} experiments to probe the interaction between gravity and these
kinds of quantum matter on large, non-Planckian distance scales, without the
necessity of performing high-energy experiments, as is required for probing
the short-range weak and strong forces on very short distance scales. \ The
quantum many-body problem, even in its nonrelativistic limit, may lead to
nontrivial interactions with weak, long-range gravitational fields. \ One is
thereby\ strongly motivated to study the interaction of quantum fluids with
weak gravity, in particular, with gravitational radiation.

One manifestation of this conceptual tension is that the way one views a
quantum fluid in QM is conceptually radically different from the way that
one views a perfect fluid in GR, where only the $local$ properties of the
fluid, which can conceptually always be spatially separated into
independent, infinitesimal fluid elements, are to be considered. \ For
example, interstellar dust particles can be thought of as being a perfect
fluid in GR, provided that we can neglect all interactions between such
particles. \ The response of these particles in this\ classical many-body
system to a gravitational wave passing over it, is characterized by the
local, free-fall motion of each dust particle. \ These particles are treated
as test particles in GR, and they do not react back upon the gravitational
wave. \ The weak equivalence principle tells us that the gravitational force
from the wave vanishes in the local, freely-falling inertial frame centered
on a given dust particle, and hence cannot do any work on this particle in
the absence of any mutual interactions between the particles. \ There is
therefore no possibility of any energy transfer at all from the
gravitational wave to this classical many-body system, so that it is
impossible to use such a system of $noninteracting$ classical particles to
construct an antenna for receiving or emitting gravitational radiation. \ 

By contrast, due to their radical delocalization, particles in a macroscopic
quantum many-body system, such as a superconductor, are entangled with each
other in such a way that there arises an unusual ``quantum rigidity'' of the
system, closely associated with what London called ``the rigidity of the
macroscopic wavefunction'' \cite{London}. \ This rigidity arises from the
single-valuedness of the macroscopic ground-state wavefunction, and the fact
that there exists an energy gap (the BCS gap) separating the ground state
from all the excited states of the system. \ There should arise from this
rigidity of the wavefunction not only the usual Meissner effect, in which
the magnetic field is expelled from the interior of the superconductor, but
also a Meissner-like effect, in which the Lense-Thirring field of the
gravitational wave is also expelled, as we shall presently see. \ 

This\ in turn should lead to a mirror-like reflection of a gravitational
plane wave from the surface of the superconductor. \ Unfortunately, there is
no way to check whether this actually occurs or not, without a practical
laboratory emitter and receiver of gravitational radiation. \ However, we
shall see that closely associated with this Meissner-like effect, there also
should arise a direct coupling between the electromagnetic (EM) and
gravitational (GR) radiation fields at the surface of a certain type of
superconductor, in a quantum kind of direct transducer action. \ In this
way, practical laboratory emitters and receivers of gravity waves could
indeed become possible. \ Thus a \textit{quantum fluid} in QM should behave
in a radically different manner from that of a \textit{perfect fluid} in GR,
in their respective responses to gravitational radiation. \ 

In an ideal BEC, there would seem to be no quantum entanglement, since the
total condensate wavefunction can be written as a tensor product of many
copies of the same one-particle wavefunction, one for each particle. \
However, as a necessary condition for the formation of any actual BEC,
including the recently discovered atomic BECs, there must be mutual
interactions between the particles, such as repulsive $S$-wave scattering. \
Each scattering event $entangles$ the momenta of the two particles that
participate in it. \ The many particle-particle scatterings, for example,
the hard-sphere collisions in superfluid helium, naturally lead to a
macroscopically entangled state. \ Moreover, in the case of
superconductivity, a given Cooper pair of electrons in the superconductor is
in a Bohm singlet state, which is an entangled state in which the two
electrons in the Cooper pair have opposite spins $and$ opposite momenta.

\subsection{Spontaneous symmetry breaking, ODLRO, and superluminality}

The unusual states of matter in these quantum fluids all possess \textit{%
spontaneous symmetry breaking}, in which the ground state, or the vacuum
state, of the quantum many-body system (any $state$, or \textit{many-body
wavefunction}, is purely a quantum concept) breaks the symmetry of the
Hamiltonian of the system (any quantum Hamiltonian, however, can be
conceived of as a classical one in the correspondence-principle limit). \
The physical vacuum, which is in an intrinsically nonlocal ground state of a
relativistic quantum field theory, possesses certain similarities to the
ground state of a superconductor. \ Weinberg has argued that in
superconductivity, the spontaneous symmetry breaking process results in a
broken $gauge$ invariance \cite{Weinberg(Nambu)}, an idea which traces back
to the early work of Nambu \cite{Nambu}. \ The Meissner effect is closely
analogous to the Higgs mechanism of high-energy physics, in which the
physical vacuum also spontaneously breaks local gauge invariance. From this
viewpoint, the appearance of the London penetration depth for a
superconductor is analogous in an inverse manner to the appearance of a mass
for a gauge boson, such as that of the W or Z boson. \ Thus, the photon,
viewed as a gauge boson, acquires a mass inside the superconductor, such
that its Compton wavelength becomes the London penetration depth. \ \qquad

Closely related to this spontaneous symmetry breaking process is the
appearance of Yang's off-diagonal long-range order (ODLRO) of the reduced
density matrix in the coordinate-space representation for these
macroscopically coherent quantum systems \cite{yang}. \ In particular, there
seems to be no limit on how far apart Cooper pairs can be inside a single
superconductor before they lose their quantum coherence. \ ODLRO and
spontaneous symmetry breaking are both purely quantum concepts. \ 

Within a superconductor there should arise both the phenomenon of
instantaneous EPR correlations-at-a-distance and the phenomenon of the
rigidity of the wavefunction, i.e., a Meissner-like response to radiation
fields. \ Both phenomena, which we shall presently see are intimately
connected, involve at the microscopic level $interactions$ of entangled
electrons with an external environment, either through local $measurements$,
such as in Bell-type experiments, or through local $perturbations$, such as
those due to gravitational radiation fields interacting locally with these
electrons. \ 

Although at first sight the notion of ``infinite quantum rigidity'' would
seem to imply infinite velocities, and hence would seem to violate
relativity, there are in fact no violations of relativistic causality here,
since the instantaneous EPR $correlations$-at-a-distance (as seen by an
observer in the center-of-mass frame) are not instantaneous $signals$%
-at-a-distance, which could instantaneously connect causes to effects \cite%
{ChiaoHeisenberg}. \ Also, experiments have verified the existence of
superluminal wave packet propagations, i.e., faster-than-$c$, infinite, and
even negative group velocities, for finite-bandwidth, analytic wave packets
in the excitations of a wide range of physical systems \cite{ChiaoSolvay}%
\cite{Superluminal}. \ An analytic function, e.g., a Gaussian wave packet,
contains sufficient information in its early tail such that a causal medium
can, during its propagation, reconstruct the entire wave packet with a
superluminal pulse advancement, and with little distortion. \ Relativistic
causality forbids\ only the $front$ velocity, i.e., the velocity of $%
discontinuities$, which connects causes to their effects, from exceeding the
speed of light $c$, but does not forbid a wave packet's $group$ velocity
from being superluminal. \ One example is the observed superluminal
tunneling of single-photon wave packets \cite{Steinberg}. \ In the case of
superconductors, the ``finite-bandwidth'' condition is guaranteed by the
restriction that the local perturbations arising from classical fields only
contain frequencies less than the BCS gap frequency, in order to satisfy the
quantum adiabatic theorem. \ Thus the notion of ``infinite quantum
rigidity,'' although counterintuitive, does not in fact violate relativistic
causality. \ I shall return to this point later in Section 3.

\subsection{The equivalence versus the uncertainty principle}

Concerning conceptual tension (2), the weak equivalence principle is
formulated at its outset using the concept of ``trajectory,'' or
equivalently, ``geodesic.'' \ By contrast, Bohr has taught us that the very $%
concept$ of trajectory must be abandoned, because of the uncertainty
principle. \ Thus the equivalence and the uncertainty principles are in a
fundamental conceptual tension. \ The equivalence principle is based on the
idea of locality, since it requires that the region of space, inside which
two trajectories of two nearby freely-falling objects of different masses,
compositions, or thermodynamic states, are to be compared, go to zero
volume, before the principle becomes exact. \ This limiting procedure is in
conflict with the uncertainty principle, since taking the limit of the
volume of space going to zero, within which these objects are to be
measured, makes their momenta infinitely uncertain. \ 

Experimental manifestations of this conceptual tension may occur not only on
microscopic, but also on macroscopic length scales, again since there exist
macroscopic quantum objects. \ However, whenever the correspondence
principle holds, the \textit{center of mass} of a quantum wavepacket (for a
single particle or for an entire quantum object) moves according to
Ehrenfest's theorem along a classical trajectory, and $then$ it is possible
to reconcile these two principles.

Davies \cite{Davies}\ has come up with a simple example of a quantum
violation of the equivalence principle \cite{onofrio}\cite{Adunas}: Consider
two perfectly elastic balls, e.g., one made out of rubber, and one made out
of steel, bouncing against a perfectly elastic table. \ If we drop the two
balls from the same height above the table, their classical trajectories,
and hence their classical periods of oscillation will be identical, and
independent of the mass or composition of the balls. \ This is a consequence
of the weak equivalence principle. \ However, quantum mechanically, there
will be the phenomenon of tunneling, in which the two balls can penetrate
into the classically forbidden region $above$ their turning points. \ The
extra time spent by the balls in the classically forbidden region due to
tunneling will depend on their mass (and thus on their composition). \ Thus
there will in principle be \textit{mass-dependent} quantum corrections of
the classical periods of the bouncing motion of these balls, which will lead
to quantum violations of the equivalence principle. \ \ 

There exist macroscopic situations in which Ehrenfest's correspondence
principle fails. \ Imagine that one is inside a large quantum object, such
as a big piece of superconconductor. \ Even in the limit of a very large
size and a very large number of particles in this object (i.e., the
thermodynamic limit), there exists no correspondence-principle limit in
which classical trajectories or geodesics for the electrons which are
members of Cooper pairs within the superconductor, make any sense. \ This
may be true in spite of the fact that the motion of its center of mass of
the superconductor may perfectly obey the equivalence principle, and
therefore may be conceptualized in terms of a geodesic. \ 

However, due to the uncertainty principle and the indistinguishability of
identical particles, the electrons of the BCS ground state of a
superconductor are radically delocalized. \ Hence one again expects that a
superconductor should respond to externally applied gravitational fields,
including radiation fields, in a radically different fashion from the
response of normal electrons undergoing geodesic motions inside a normal
metal. \ Experimentally, this radical difference should show up when one
cools a normal metal through the superconducting transition temperature, so
that it becomes a superconductor.

\section{Superconductors as antennas for gravitational radiation\ }

Another\ strong motivation for performing the quantum calculation to be
given below, is that it predicts a large, counterintuitive \textit{quantum
rigidity }of a macroscopic wavefunction, such as that in a big piece of
superconductor, when it interacts with externally applied gravitational\
radiation fields \cite{ChiaoPRB}. \ Mathematically, this quantum rigidity
corresponds to the statement that the macroscopic wavefunction of the
superconductor must remain \textit{single-valued} at all times during the
changes arising from adiabatic perturbations due to radiation fields. \ This
implies that objects such as superconductors should be much better
gravitational-wave antennas than Weber bars. \ 

The rigidity of the macroscopic wavefunction\ of the superconductor
originates from the instantaneous Einstein-Podolsky-Rosen (EPR)
correlations-at-a-distance in the behavior of a Cooper pair of electrons in
the Bardeen-Cooper-Schrieffer (BCS) ground state in distant parts of the
superconductor viewed as a single quantum system, when there exists some
kind of gap, such as the BCS gap, which keeps the entire system
adiabatically in its ground state during perturbations due to radiation. \
Two electrons which are members of a single Cooper pair are in a Bohm
singlet state, and hence are quantum-mechanically entangled with each other,
in the sense that they are in a superposition state of opposite spins $and$
opposite momenta. \ This quantum entanglement gives rise to EPR correlations
at long distance scales within the superconductor. \ The electrons in a
superconductor in its ground BCS state are not only $macroscopically$
entangled, but due to the existence of the BCS gap which separates the BCS
ground state energetically from all excited states, they are also $%
protectively$ entangled, in the sense that this entangled state is protected
by the presence of the BCS gap from decoherence arising from the thermal
environment, provided that the system temperature is kept well below the BCS
transition temperature. \ 

The resulting large quantum rigidity is in contrast to the tiny rigidity of
classical matter, such as that of the normal metals used in Weber bars, in
their response to gravitational radiation. \ The essential difference
between quantum and classical matter is that there can exist macroscopic
quantum interference, and hence macroscopic quantum coherence, throughout
the entire quantum system, which is absent in a classical system.\ \ 

One manifestation of the tiny rigidity of classical matter is the fact that
the speed of sound in a Weber bar is typically five orders of magnitude less
than the speed of light. \ In order to transfer energy coherently from a
gravitational wave by classical means, for example, by acoustical modes
inside the bar\ to some local detector, e.g., a piezoelectric crystal glued
to the middle of the bar, the length scale of the Weber bar $L$ is limited
to a distance scale on the order of the speed of sound times the period of
the gravitational wave, i.e., an acoustical wavelength $\lambda _{sound}$,
which is typically five orders of magnitude smaller than the gravitational
radiation wavelength $\lambda $ to be detected. \ This makes the Weber bar,
which is thereby limited in its length to $L\simeq \lambda _{sound}$, much
too short an antenna to couple efficiently to free space. \ 

However, macroscopic quantum objects such as superconductors used as
antennas are not limited by these classical considerations, but can have a
length scale $L$ on the same order as (or even much greater than) the
gravitational radiation wavelength $\lambda $. \ Since the radiation
efficiency of a quadrupole antenna scales as the length of the antenna $L$
to the fourth power when $L<<\lambda $, quantum antennas should be much more
efficient in coupling to free space than classical ones like the Weber bar
by at least a factor of $\left( \lambda /\lambda _{sound}\right) ^{4}$. \
Also, we shall see below that a certain type of superconductor may be a $%
transducer$ for directly converting gravitational waves into electromagnetic
waves, and vice versa; this then dispenses altogether with the necessity of
the use of piezoelectric crystals as transducers.

Weinberg \cite{Weinberg}\ gives a measure of the efficiency of coupling of a
Weber bar antenna of mass $M$, length $L$, and velocity of sound $v_{sound}$%
, in terms of a branching ratio for the emission of gravitational radiation
by the Weber bar, relative to the emission of heat, i.e., the ratio of the $%
rate$ of emission of gravitational radiation $\Gamma _{grav}$ relative to
the $rate$ of the decay of the acoustical oscillations into heat $\Gamma
_{heat}$, which is given by%
\begin{equation}
\eta \equiv \frac{\Gamma _{grav}}{\Gamma _{heat}}=\frac{64GMv_{sound}^{4}}{%
15L^{2}c^{5}\Gamma _{heat}}\simeq 3\times 10^{-34}.  \label{WeinbergEff}
\end{equation}%
The quartic power dependence of the efficiency $\eta$\ on the velocity of
sound $v_{sound}$ arises from the quartic dependence of the coupling
efficiency to free space of a quadrupole antenna upon its length $L$, when $%
L<<\lambda $. \ 

Assuming for the moment that the quantum rigidity of a superconductor allows
us to replace the velocity of sound $v_{sound}$ by the speed of light $c$
(i.e., that the typical size $L$ of a quantum antenna bar can become as
large as the wavelength $\lambda $), we see that superconductors can be more
efficient than Weber bars, based on the $v_{sound}^{4}$ factor alone, by
twenty orders of magnitude, i.e., 
\begin{equation}
\left( \frac{c}{v_{sound}}\right) ^{4}\simeq 10^{20}.
\end{equation}%
Thus, even if it should turn out that superconducting antennas in the final
analysis are still not very efficient $generators$ of gravitational
radiation, they should be much more efficient $receivers$ of this radiation
than Weber bars for detecting astrophysical sources of gravitational
radiation \cite{AnandanChiao}\cite{anandan1985}\cite{PengTorr}. \ However, I
shall give arguments below as to why under certain circumstances involving
``natural impedance matching'' between quadrupolar EM and GR plane waves
upon a mirror-like reflection at the planar surface of extreme type II,
dissipationless superconductors, the efficiency of such superconductors used
as simultaneous transducers and antennas for gravitational radiation, might
in fact become of the order of unity, so that a gravitational analog of
Hertz's experiment might then become possible.\qquad

But why should the speed of sound in superconductors, which is not much
different from that in normal metals, not also characterize the rigidity of
a superconducting metal when it is in a superconducting state? \ What is it
about a superconductor in its superconducting state that makes it so
radically different from the same metal when it is in a normal state? \ The
answer lies in the important distinction between $longitudinal$ and $%
transverse$ mechanical excitations of the superconductor \cite%
{CurrentCorrelation}. \ Whereas longitudinal, compressional sound waves
propagate in superconductors at normal sound speeds, transverse excitations,
such as those\ induced by a gravitational plane wave incident normally on a
slab of superconductor in its superconducting state, cannot so propagate. \ 

Suppose that the opposite were true, i.e., that there were $no$ substantial
difference between the response of a superconductor and a normal metal to
gravitational radiation. \ (Let us assume for the moment the complete
absence of any electromagnetic radiation.) \ Then the interaction of the
superconductor with gravitational radiation, either in its normal or in its
superconducting state, will be completely negligible, as is indicated by Eq.(%
\ref{WeinbergEff}). \ We would then expect the gravitational wave to
penetrate deeply into the interior of the superconductor (see Figure \ref%
{circulations}).

If this were true, the motion of a Cooper pair deep inside the
superconductor (i.e., deep on the scale of the London penetration depth),
could then be characterized by a velocity $\mathbf{v}_{pair}(t)$ which would
not be appreciably different from the velocity of a normal electron or of a
nearby lattice ion (we shall neglect the velocity of sound in this argument,
since $v_{sound}<<c$). \ Hence locally, by the weak equivalence principle,
Cooper pairs, normal electrons, and lattice ions (i.e., independent of the
charges and masses of these particles) would all undergo free fall together,
so that 
\begin{equation}
\mathbf{v}_{pair}(t)=-\mathbf{h}(t),
\end{equation}%
where $\mathbf{v}_{pair}(t)$ is the local velocity of a Cooper pair, and
where $-\mathbf{h}(t)$ is the local velocity of a classical test particle,
whose motion is induced by the presence of the gravitational wave, as seen
by an observer sitting in an inertial frame located at the center of mass of
the superconductor. Then the $curl$ of the velocity field $\mathbf{v}%
_{pair}(t)$\ deep inside the superconductor, as seen by this observer, would
be nonvanishing%
\begin{equation}
\mathbf{\nabla \times v}_{pair}(t)=-\mathbf{\nabla \times h}(t)=-\mathbf{B}%
_{G}(t)\neq 0,
\end{equation}%
since the Lense-Thirring or gravitomagnetic field $\mathbf{B}_{G}(t)$ of
gravitational radiation does not vanish deep in the interior of the
superconductor when gravitational radiation is present (we shall see
presently that the $\mathbf{h}(t)$ field plays the role of a gravitomagnetic
vector potential, just like the usual vector potential $\mathbf{A}(t)$ in
the electromagnetic case). \ 

\begin{figure}[tbp]
\centerline{\includegraphics{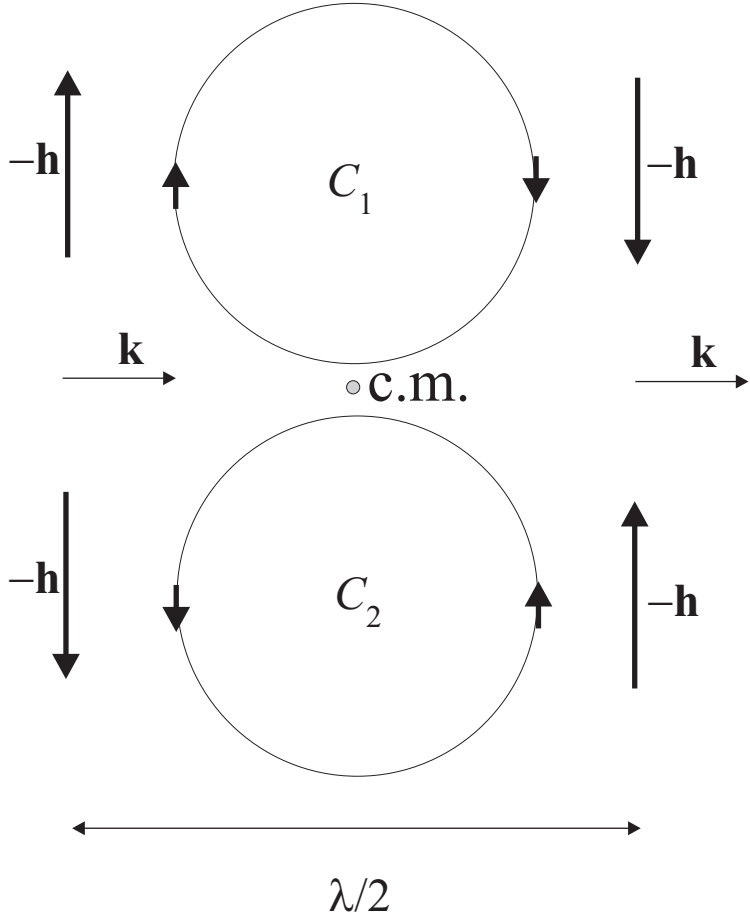}}
\caption{A side-view snapshot of a monochromatic gravitational plane wave
inside a thick superconducting slab, propagating to the right with a wave
vector $\mathbf{k}$ along the $z$ axis, which induces tidal motions along
the $x$ axis resulting in the velocity field $\mathbf{-h}$ of test
particles, as seen by an observer sitting in an inertial frame centered on
the center-of-mass (c.m.) of the superconductor. After half a wavelength of
propagation, these tidal motions reverse sign. Hence there exists
nonvanishing circulations around the circuits $C_{1}$ and $C_{2}$, i.e., $%
fluxes$ of the gravitomagnetic (or Lense-Thirring) field inside $C_{1}$ and $%
C_{2}$. These fluxes are directly proportional to the quantities given by
Eqs.(\ref{Stokes}) and (\ref{TwinParadox}), and are gauge-invariant. For
small circuits, they are also directly proportional to the nonvanishing
Riemann curvature tensor component $R_{\widehat{0}\widehat{x}\widehat{z}%
\widehat{x}}$. The propagation depicted here of a GR wave penetrating deeply
into the interior of a thick slab of superconductor (which is thick compared
to the wavelength $\protect\lambda $) would violate the single-valuedness of
the wavefunction. However, the existence of a Meissner-like effect, in which
all the radiation fields of the GR wave, including its Lense-Thirring
fields, are totally expelled from the superconductor (apart from a thin
surface layer of the order of the London penetration depth), would prevent
such a violation.}
\label{circulations}
\end{figure}

This, however, leads to a contradiction. \ It is well known that for
adiabatic perturbations (e.g., for gravity waves whose frequencies are
sufficiently far below the BCS gap frequency, so that the entire quantum
system remains $adiabatically$ in its ground state), the superfluid velocity
field $\mathbf{v}_{pair}(t)$ deep in the interior of the superconductor
(i.e., at a depth much greater than the London penetration depth) must
remain $irrotational$ at all times \cite{putterman}, i.e.,%
\begin{equation}
\mathbf{\nabla \times v}_{pair}(t)=0.
\end{equation}%
Otherwise, if this irrotational condition were not satisfied in the presence
of gravitational radiation, the wavefunction would not remain single-valued.
\ Deep inside the superconductor, $\mathbf{v}_{pair}(t)=\frac{\hbar }{m_{2}}%
\mathbf{\nabla }\phi (t)$, where $m_{2}$ is the mass of the Cooper pair. \
Thus the superfluid or Cooper pair velocity is directly proportional to the
spatial gradient of the phase $\phi (t)$ of the Cooper-pair condensate
wavefunction. \ It would then follow, from such a\ supposed violation of the
irrotational condition, that the phase shift $\Delta \phi (t)$ after one
round-trip back to the same point around an arbitrary closed curve $C$ deep
inside the superconductor in its ground state, is, upon using Stokes's
theorem, 
\begin{eqnarray}
\Delta \phi (t) &=&\oint_{C}\mathbf{\nabla }\phi (t)\cdot d\mathbf{l}=\frac{%
m_{2}}{\hbar }\oint_{C}\mathbf{v}_{pair}(t)\cdot d\mathbf{l}  \notag \\
&=&\frac{m_{2}}{\hbar }\iint_{S(C)}\mathbf{\nabla \times v}_{pair}(t)\cdot d%
\mathbf{S=-}\frac{m_{2}}{\hbar }\iint_{S(C)}\mathbf{B}_{G}(t)\cdot d\mathbf{S%
}\neq 0,  \label{Stokes}
\end{eqnarray}%
where $S(C)$ is a surface bounded by $C$ (see Figure \ref{circulations}). \
That is, the phase anomaly $\Delta \phi (t)$ of the wavefunction is directly
proportional to the flux through the circuit $C$ of the gravitomagnetic
field $\mathbf{B}_{G}(t)$, which is nonvanishing. \ This phase anomaly $%
\Delta \phi (t)$ is also directly proportional to a \textit{nonvanishing}
component of the Riemann curvature tensor. \ However, that the round-trip
phase shift $\Delta \phi (t)$ is nonvanishing in the ground state
wavefunction of any quantum system, is impossible in QM due to the \textit{%
single-valuedness} of the wavefunction, as viewed from the inertial frame of
the center-of-mass of the system \cite{Abrikosov}. \ Otherwise, the very
concept of \textit{wavefunction} would become meaningless.

There is extensive experimental evidence that the single-valuedness of the
wavefunction in QM is not violated. \ For example, the observed quantization
of orbital angular momentum in atoms and molecules constitutes such evidence
on microscopic length scales. \ Also, the observations of quantization of
the circulation of vortices in both superfluids helium of isotope 3 and
helium of isotope 4, and of the quantization of flux in superconductors,
constitute such evidence on macroscopic length scales. \ As a special case
of the latter when the topological winding number is zero, the Meissner
effect is itself evidence for the validity of the single-valuedness of the
macroscopic wavefunction. \ 

Therefore, in the presence of a gravity wave, Cooper pairs $cannot$ undergo
free fall, and the $transverse$ excitations of the Cooper pair condensate
must remain rigidly irrotational at all times in the adiabatic limit. \ This
leads to a Meissner-like effect in which the Lense-Thirring field $\mathbf{B}%
_{G}(t)$ is expelled, and would seemingly lead to infinite velocities for
the transverse excitations inside the superconductor. \ (See the above
discussion of superluminality.) \ However, such transverse excitations
should be coupled\ to\ perturbations of spacetime through the Maxwell-like
equations for the time-varying gravitational fields to be discussed below,
and then the speed of such excitations may turn out to be governed by the
vacuum speed of light $c$. \ \qquad\ \ 

\section{Meissner-like effect in the response of a superconductor to
gravitational radiation}

The calculation for the\ $quantum$ response of large objects, for example, a
big piece of superconductor, to weak gravitational radiation, is based on
the concept of $wavefunction$, or \textit{quantum state}, for example, the
BCS state, and proceeds along completely different lines from the
calculation for the $classical$ response of a Weber bar to this radiation,
which is based on the concept of $geodesic$, or \textit{classical trajectory}
\cite{MTW}. \ When the frequency of the gravitational radiation is\ much
less than the BCS gap, the entire superconductor should evolve in time in
accordance with the quantum adiabatic theorem, and should therefore stay
rigidly, i.e., adiabatically, in its ground state. There results a large,
diamagnetic-like \textit{linear response} of the entire superconductor to
externally applied, time-varying gravitational fields. \ This Meissner-like
effect does not alter the geodesic center-of-mass motion of the
superconductor, but radically alters the internal behavior of its electrons,
which are all radically delocalized due to entanglement within the
superconductor.

Here I would like to give a brief historical survey of this problem. \ In
1966, DeWitt \cite{DeWitt}\ considered the interaction of a superconductor
with gravitational fields, in particular with the Lense-Thirring field. \
Starting from the general relativistic Lagrangian for a single electron with
a charge $e$ and a mass $m$, he derived in the limit of weak gravity and
slow particles a nonrelativistic Hamiltonian for a single electron in the
superconductor, which satisfied the minimal-coupling rule%
\begin{equation}
\mathbf{p\rightarrow p}-e\mathbf{A}-m\mathbf{h},
\end{equation}%
where $\mathbf{p}$ is the canonical momentum, $\mathbf{A}$ is the usual
vector potential, and $\mathbf{h}$ is a gauge-like vector potential formed
from the three space-time components $g_{i0}$ of the metric tensor viewed as
an ordinary three-vector. \ Papini \cite{Papini} in 1967 considered the
possibility of the detection the quantum phase shift induced by $\mathbf{h}$
arising from the Lense-Thirring field generated by a nearby rotating massive
body, by means of a superconducting interference device (or SQUID) using
Josephson junctions.\ \ In 1979, Stodolsky \cite{Stodolsky}\ considered the
detection of phase shifts by means of matter-wave and light-wave
interferometry (e.g., neutron interferometers, such as that of Collela,
Overhauser, and Werner \cite{COW}, and light-wave interferometers, such as
that of LIGO (i.e, the Laser-Interferometer Gravitational-wave Observatory;
e.g., see the review by Tyson and Giffard \cite{TysonGiffard})), in the
approximation of a semiclassical, single-particle propagation in weak
gravitational fields, such as those associated with gravity-wave and
Lense-Thirring fields. \ In 1983, Ross \cite{Ross} derived the modified
London equations for a superconductor in a gravitational field, and showed
that these equations are consistent with the modified fluxoid quantization
condition in a gravitational field found earlier by DeWitt in 1966.

In a series of papers in the early 1980s, Anandan and I considered the
possibility of constructing antennas for detecting time-varying
Lense-Thirring fields, and thus for detecting gravitational radiation, using
Josephson junctions as transducers, in $neutral$ superfluid helium analogs
of the SQUID using an antenna geometry in the form of a figure 8 superfluid
loop, and also an antenna bent into a the form of a baseball seam \cite%
{AnandanChiao}.\ \ In 1985, Anandan \cite{anandan1985}\ considered the
possibility of using superconducting circuits as detectors for astrophysical
sources of gravitational radiation, but did not mention the possibility of
superconductors being efficient emitters, and thus practical laboratory
sources of gravity waves, as is being considered here. \ In 1990, Peng and
Torr used the generalized London equations to treat the interaction of a
bulk superconductor with gravitational radiation, and concluded that such a
superconducting antenna would be many orders of magnitude more sensitive
than a Weber bar \cite{PengTorr}. \ There have also been earlier predictions
of a modified Meissner effect in the response of superconductors to
time-varying Lense-Thirring fields, and hence to gravitational radiation %
\cite{Li1991}\cite{Li1992}. \ For recent work along these lines, see \cite%
{Fischer}. \ These papers, however, also did not consider the possibility of
a transducer action between EM and GR radiation mediated by the
superconductor, as is being considered here. \ Also, the theoretical
approach\ taken here is quite different, as our approach will be based on
the Ginzburg-Landau theory of superconductivity, and the resulting
constitutive relation for the gravitomagnetic or Lense-Thirring field,
rather than on the modified London equations.\ 

I shall show below that Josephson junctions, which are difficult to
implement experimentally, are unnecessary, and that a superconductor can by
itself be a $direct$ transducer from electromagnetic to gravitational
radiation upon reflection of the wave from a vacuum-superconductor
interface, with a surprisingly good conversion efficiency. \ By reciprocity,
this conversion process can be reversed, so that gravitational radiation can
also be converted upon reflection into electromagnetic radiation from the
same interface, with equal efficiency. \ The geometry of a superconducting
slab-shaped antenna proposed here is much simpler than some of the earlier
proposed antenna geometries. \ These developments suggest the possibility of
a simple, Hertz-like experiment, in which the emission and the reception of
gravitational radiation at microwave frequencies can be implemented by means
of a pair of superconductors used as transducers. \ 

\subsection{Calculation of diamagnetic-like coupling energies: The
interaction Hamiltonian}

Consider a gravitational plane wave propagating along the $z$ axis, which
impinges at normal incidence on a piece of superconductor in the form of a
large circular slab of radius $r_{0}$ and of thickness $d$. \ Let the radius 
$r_{0}$ be much larger than the wavelength $\lambda $ of the plane wave, so
that one can neglect diffraction effects. \ Similarly, let $d$ be much
thicker than $\lambda $. \ For simplicity, let the superconductor be at a
temperature of absolute zero, so that only quantum effects need to be
considered. \ 

The calculation of the coupling energy of the superconductor in the
simultaneous presence of both electromagnetic and gravitational fields
starts from the Lagrangian for a single particle of rest mass $m$ and charge 
$e$ (i.e., an electron, but neglecting its spin)%
\begin{equation}
L=-m(-g_{\mu \nu }\dot{x}^{\mu }\dot{x}^{\nu })^{1/2}+eA_{\mu }\dot{x}^{\mu
},
\end{equation}%
from which a minimal-coupling form of the Hamiltonian\ for an electron in a
superconductor, in the limit of $weak$ gravitational fields and $low$
velocities, has been derived by DeWitt \cite{DeWitt}. \ Here we apply this
minimal-coupling Hamiltonian to $pairs$ of electrons (i.e, Cooper pairs in
spin-zero singlet states),%
\begin{equation}
H=\frac{1}{2m_{2eff}}\left( \mathbf{p}-e_{2}\mathbf{A}-m_{2}\mathbf{h}%
\right) ^{2}  \label{DeWittHamiltonian}
\end{equation}%
(using SI units), where $m_{2}=2m_{e}$ is the vacuum rest mass of a Cooper
pair, $m_{2eff}$\ is its effective mass, $e_{2}=2e\ $is its charge, $\mathbf{%
p}$\ is its canonical momentum, $\mathbf{A}$\ is the electromagnetic vector
potential, and $\mathbf{h}$\ is the gravitomagnetic vector potential, which
is the gravitational analog of $\mathbf{A}$ in the case of weak gravity. \
The gravitomagnetic vector potential $\mathbf{h}$\ is the three-velocity
formed from the space-time components $h_{i0}$ of the small deviations of
the metric tensor $h_{\mu \nu }=g_{\mu \nu }-\eta _{\mu \nu }$ from flat
spacetime (the metric tensor being given by $g_{\mu \nu }$, and the
Minkowski tensor for flat spacetime being given by $\eta _{\mu \nu }=\mathrm{%
diag}(-1,1,1,1)$). \ Thus we shall define%
\begin{equation}
\left. \mathbf{h}\right| _{i}\equiv h_{i0}c\text{ .}  \label{hVector}
\end{equation}%
It is convenient for performing this calculation to choose the radiation
gauge for both $\mathbf{A}$\ and\ $\mathbf{h}$, so that%
\begin{equation}
\nabla \cdot \mathbf{A}=\nabla \cdot \mathbf{h}=0\mathbf{,}
\label{RadiationGauge}
\end{equation}%
where the chosen coordinate system is that of an inertial frame which
coincides with the freely-falling center of mass of the superconductor at
the origin (this is $not$ the transverse-traceless gauge choice). \ The
physical meaning of\ $\mathbf{h}$\ is that, apart from a sign, it is the
three-velocity of a local, freely-falling test particle as seen by an
observer in an inertial frame located at the center of mass of the
superconductor. \ In Eq.(\ref{DeWittHamiltonian}), we have neglected for the
moment the interactions of the Cooper pairs with each other.

Why not use the standard transverse-traceless gauge in order to perform
these calculations? \ The answer is given in Figure \ref{circulations}, in
which we depict a side-view snapshot of a gravitational plane wave
propagating to the right. \ The arrows indicate the instantaneous velocity
vectors $-\mathbf{h}$\ of the test particles induced by the wave, as seen by
an inertial observer at the center-of-mass. \ Note that the gravitational
tidal forces reverse in sign after a propagation by half a wavelength to the
right along the $\mathbf{k}$ axis. \ Therefore, by inspection\ of the
diagram, we see that the circulation integrals around circuits $C_{1}$ and $%
C_{2}$ 
\begin{equation}
\oint_{C_{1}}\mathbf{h}\cdot d\mathbf{l}\neq 0,\text{and }\oint_{C_{2}}%
\mathbf{h}\cdot d\mathbf{l}\neq 0  \label{circulation}
\end{equation}%
do not vanish. \ These circulation integrals are gauge-invariant quantities,
since there are related to the gauge-invariant general relativistic time
shift $\Delta t$ (and the corresponding quantum phase shift), where, for
weak gravity and small circuits $C$,%
\begin{equation}
\Delta t=-\oint_{C}\frac{g_{0i}dx^{i}}{g_{00}}\approx \frac{1}{c}\oint_{C}%
\mathbf{h}\cdot d\mathbf{l}\neq 0.  \label{TwinParadox}
\end{equation}%
The time shift $\Delta t$ is related through Stokes's theorem (see Eq.(\ref%
{Stokes})) to the $flux$ of the gravitomagnetic (or Lense-Thirring) field
through the circuit $C$, which, for small circuits $C$ close to the center
of mass, is directly proportional to the \textit{nonvanishing} ``magnetic''
Riemann curvature tensor component $R_{\widehat{0}\widehat{x}\widehat{z}%
\widehat{x}}$, where the hats above the indices indicate the use of Fermi
normal coordinates \cite{MisnerManasse}.

Since in the transverse-traceless gauge, $\mathbf{h}$ is chosen to be
identically zero, there would be no way to satisfy Eqs.(\ref{circulation})
and (\ref{TwinParadox}). \ In the long-wavelength limit, i.e., in the case
where the antenna is small compared to a wavelength, such as in Weber bars,
the transverse-traceless gauge can be a valid and more convenient choice
than the radiation gauge being used here. \ However, we wish to\ be able to
consider the case of superconducting slabs which are large compared to a
wavelength, where the long-wavelength approximation breaks down, and
therefore we cannot use the transverse-traceless gauge, but must use the
radiation gauge instead.

The electromagnetic vector\textbf{\ }potential $\mathbf{A}$\ in the above
minimal-coupling Hamiltonian gives rise to Aharonov-Bohm interference. \ In
like manner, the gravitomagnetic vector potential $\mathbf{h}$ gives rise to
a general relativistic twin ``paradox'' for rotating coordinate systems and
for Lense-Thirring fields given by Eq.(\ref{TwinParadox}). \ Therefore $%
\mathbf{h}$ gives rise to Sagnac interference in both light and matter
waves. \ The Sagnac effect has recently been observed in superfluid helium
interferometers using Josephson junctions, and has been used to detect the
Earth's rotation around its polar axis \cite{Simmonds}.

From the above Hamiltonian, we see that the minimal coupling rule for Cooper
pairs now becomes%
\begin{equation}
\mathbf{p\rightarrow p}-e_{2}\mathbf{A}-m_{2}\mathbf{h}\text{ }
\label{Minimal coupling rule}
\end{equation}%
in the simultaneous presence of electromagnetic (EM) and weak general
relativistic (GR) fields. \ This minimal-coupling rule has been
experimentally tested in the case of a uniformly rotating superconducting
ring, since it predicts the existence of a London magnetic moment for the
rotating superconductor, in which magnetic flux is generated through the
center of the ring due to its rotational motion with respect to the local
inertial frame. \ The consequences of the above minimal coupling rule for
the slightly different geometry of a uniformly rotating superconducting
sphere can be easily worked out as follows: Due to the single-valuedness of
the wavefunction, the Aharonov-Bohm and Sagnac phase shifts deep inside the
superconducting sphere (i.e., in the interior far away from the surface)
arising from the $\mathbf{A}$ and the $\mathbf{h}$ terms, must cancel each
other exactly. \ Thus the minimal coupling rule leads to a relationship
between the $\mathbf{A}$ and the $\mathbf{h}$ fields inside the bulk given
by 
\begin{equation}
e_{2}\mathbf{A}=-m_{2}\mathbf{h}\text{ .}  \label{A and h}
\end{equation}%
This relationship in turn implies that\ a uniform magnetic field $\mathbf{%
B=\nabla \times A}$, where $\mathbf{A=}\frac{1}{2}\mathbf{B\times r}$ in the
symmetric gauge, will be generated in the interior of the superconducting
sphere due to its uniform rotational motion at an angular velocity $\mathbf{%
\Omega }$ with respect to the local inertial frame, where $\mathbf{h=\Omega
\times r}$ in the rotating frame. \ \ Thus the London moment effect will
manifest itself here as a uniform magnetic field $\mathbf{B}$ in the
interior of the rigidly rotating sphere, which can be calculated by taking
the curl of both sides of\ Eq.(\ref{A and h}), and yields%
\begin{equation}
\mathbf{B=-}\frac{2m_{2}}{e_{2}}\mathbf{\Omega }\text{ ,}  \label{Larmor}
\end{equation}%
which is consistent with Larmor's theorem. \ In general, the proportionality
constant of the London moment will be given by the inverse of the
charge-to-mass ratio $e_{2}/m_{2}$, where $m_{2}$ has been experimentally
determined to be the $vacuum$ value of the Cooper pair rest mass, apart from
a small discrepancy of the order of ten parts per million, which has not yet
been completely understood \cite{Cabrera}. \ 

However, in the above argument, we have been assuming rigid-body rotation
for the entire body of the superconductor, which is obviously not valid for
microwave-frequency gravitational radiation fields, since the lattice cannot
respond to such high frequencies in such a rigid manner. Hence the above
analysis applies\ only to time-independent (i.e., magnetostatic) and
spatially homogeneous (i.e., uniform) magnetic fields and steady rotations,
and is not valid for the high-frequency, time-dependent, and spatially
inhomogeneous case of the interaction of gravitational and electromagnetic
radiation fields near the surface of the superconductor, since the above
magnetostatic analysis ignores the important boundary-value and
impedance-matching problems for radiation fields at the
vacuum-superconductor interface, which will be considered below.

One can generalize the above time-independent minimal-coupling Hamiltonian
to adiabatic time-varying situations as follows:%
\begin{equation}
H=\frac{1}{2m_{2eff}}\left( \mathbf{p}-e_{2}\mathbf{A}(t)-m_{2}\mathbf{h}%
(t)\right) ^{2}\text{ ,}  \label{TimeDependentHamiltonian}
\end{equation}%
where $\mathbf{A}(t)$ and $\mathbf{h}(t)$ are the vector potentials
associated with low-frequency electromagnetic and gravitational radiation
fields, for example. \ (This time-dependent Hamiltonian can also of course
describe low-frequency time-varying tidal and Lense-Thirring fields, as well
as radiation fields, but the adiabatic approximation can still be valid for
radiation fields oscillating at high microwave frequencies, since the BCS
gap frequency of many superconductors lie in the far-infrared part of the
spectrum.) \ Again, it is natural to choose to use the radiation gauge for
both $\mathbf{A}(t)$ and $\mathbf{h}(t)$ vector potentials in a symmetrical
manner, in the description of these time-varying fields. \ The physical
meaning of\ $\mathbf{h}(x,y,z,t)\equiv \mathbf{h}(t)$\ is that it is the $%
negative$ of the time-varying three-velocity field $\mathbf{v}%
_{test}(x,y,z,t)$ of a system of noninteracting, locally freely-falling
classical test particles as seen by the observer sitting in an inertial
frame located at the center of mass of the superconductor. \ At first, we
shall treat both $\mathbf{A}(t)$ and $\mathbf{h}(t)$ as classical fields,
but shall treat the matter, i.e., the superconductor, quantum mechanically,
in the standard semiclassical approximation. \ 

The time-dependent Hamiltonian given by Eq.(\ref{TimeDependentHamiltonian})
is, I stress, only a ``guessed'' form of the Hamiltonian, whose ultimate
justification must be an experimental one. \ In case of the time-dependent
vector potential $\mathbf{A}(t)$, there have already been many experiments
which have justified this ``guess,'' but there have been no experiments
which have tested the new term involving $\mathbf{h}(t)$. \ However, one
justification for this new term is that in the static limit, this
``guessed'' Hamiltonian goes over naturally to the magnetostatic
minimal-coupling form, which, as we have seen above, $has$ been tested
experimentally.

From Eq.(\ref{TimeDependentHamiltonian}), we see that the time-dependent
generalization of the minimal-coupling rule for Cooper pairs is%
\begin{equation}
\mathbf{p\rightarrow p}-e_{2}\mathbf{A}(t)-m_{2}\mathbf{h}(t).
\label{t-dependent minimal coupling}
\end{equation}%
It would be hard to believe that one is allowed to generalize $\mathbf{A}$
to $\mathbf{A}(t)$, but that somehow one is $not$ allowed to generalize $%
\mathbf{h}$ to $\mathbf{h}(t)$.

One important consequence that follows immediately from expanding the square
in Eq.(\ref{TimeDependentHamiltonian}) is that there exists a cross-term %
\cite{Solli}%
\begin{equation}
H_{int}=\frac{1}{2m_{2eff}}\left\{ 2e_{2}m_{2}\mathbf{A}(t)\cdot \mathbf{h}%
(t)\right\} =\left( \frac{m_{2}}{m_{2eff}}\right) e_{2}\mathbf{A}(t)\cdot 
\mathbf{h}(t).  \label{Hint}
\end{equation}%
It should be emphasized that Newton's constant $G$ does not enter here. \
The physical meaning of this interaction Hamiltonian $H_{int}$ is that there
should exist a $direct$ coupling between electromagnetic and gravitational
radiation mediated by the superconductor that involves the charge\ $e_{2}$,
and not $G$, as its coupling constant, in the quantum adiabatic theorem
limit. \ Thus the strength of this coupling is electromagnetic, and not
gravitational, in its character. \ Furthermore, the $\mathbf{A\cdot h}$\
form of $H_{int}$ implies that there should exist a \textit{linear and
reciprocal} coupling between these two radiation fields mediated by the
superconductor. \ This implies that the superconductor should be a \textit{%
quantum-mechanical\ transducer} between these two forms of radiation, which
can, in principle, convert power from one form of radiation into the other,
and vice versa, with equal efficiency.

We can see more clearly the significance of the interaction Hamiltonian $%
H_{int}$ once we convert it into second quantized form and express it in
terms of the creation and annihilation operators for the positive frequency
parts of the\ two radiation fields, as in the theory of quantum optics, so
that in the rotating-wave approximation%
\begin{equation}
H_{int}\propto a^{\dagger }b+b^{\dagger }a
\end{equation}%
where the annihilation operator $a$ and the creation operator $a^{\dagger }$
of a single classical mode of the electromagnetic radiation field, obey the
commutation relation $[a,a^{\dagger }]=1$, and where the annihilation
operator $b$ and the creation operator $b^{\dagger }$ of a matched single
classical mode of the gravitational radiation field, obey the commutation
relation $[b,b^{\dagger }]=1$. \ (This represents a crude, first attempt at
quantizing the gravitational field, which applies only in the case of weak
gravity.) \ The first term $a^{\dagger }b$ then corresponds to the process
in which a graviton is annihilated and a photon is created inside the
superconductor, and similarly the second term $b^{\dagger }a$ corresponds to
the reciprocal process, in which a photon is annihilated and a graviton is
created inside the superconductor. \ Energy is conserved by both of these
processes. \ Time-reversal symmetry, and hence reciprocity, is also
respected by this interaction Hamiltonian.

\subsection{Calculation of diamagnetic-like coupling energies: The
macroscopic wavefunction}

At this point, we need to introduce the purely quantum concept of\textit{\
wavefunction}, in conjunction with the quantum adiabatic theorem. \ To
obtain the response of the superconductor, we must make explicit use of the
fact that the ground state wavefunction of the system is $globally$
unchanged (i.e., ``rigid'') during the time variations of\ both $\mathbf{A}%
(t)$ and $\mathbf{h}(t)$. \ The condition for validity of the quantum
adiabatic theorem here is that the frequency of the perturbations $\mathbf{A}%
(t)$ and $\mathbf{h}(t)$ must be low enough compared with the BCS gap
frequency of the superconductor, so that no transitions are permitted out of
the BCS ground state of the system into any of the excited states of the
system. \ However, ``low enough'' can, in practice, still mean quite high
frequencies, e.g., microwave frequencies in the case of high $T_{c}$
superconductors, so that it becomes practical for the superconductor to
become comparable in size to the microwave wavelength $\lambda $.

Using the quantum adiabatic theorem, one obtains in first-order perturbation
theory the coupling energy $\Delta E_{int}^{(1)}$ of the superconductor in
the simultaneous presence of both $\mathbf{A}(t)$ and $\mathbf{h}(t)$
fields, which is given by%
\begin{equation*}
\Delta E_{int}^{(1)}=\left( \frac{m_{2}}{m_{2eff}}\right) \left\langle \psi
\left| e_{2}\mathbf{A}(t)\cdot \mathbf{h}(t)\right| \psi \right\rangle =
\end{equation*}%
\begin{equation}
\left( \frac{m_{2}}{m_{2eff}}\right) \int_{V}dxdydz\text{ }\psi ^{\ast
}(x,y,z)\mathbf{A}(x,y,z,t)\cdot \mathbf{h}(x,y,z,t)\psi (x,y,z)
\end{equation}%
where%
\begin{equation}
\psi (x,y,z)=\left( N/\pi r_{0}^{2}d\right) ^{1/2}=\text{Constant}
\end{equation}%
is the Cooper-pair condensate wavefunction (or Ginzburg-Landau order
parameter) of a homogeneous superconductor of volume $V$ \cite{homogeneous},
the normalization condition having been imposed that%
\begin{equation}
\int_{V}dxdydz\text{ }\psi ^{\ast }(x,y,z)\psi (x,y,z)=N,
\end{equation}%
where $N$ is the total number of Cooper pairs in the superconductor. \
Assuming that both $\mathbf{A}(t)$ and $\mathbf{h}(t)$ have the same (``+'')
polarization of quadrupolar radiation, and that both plane waves impinge on
the slab of superconductor at normal incidence, then in Cartesian
coordinates,%
\begin{equation}
\mathbf{A}(t)=(A_{1}(t),A_{2}(t),A_{3}(t))=\frac{1}{2}(x,-y,0)A_{+}\cos
(kz-\omega t)
\end{equation}%
\begin{equation}
\mathbf{h}(t)=(h_{1}(t),h_{2}(t),h_{3}(t))=\frac{1}{2}(x,-y,0)h_{+}\cos
(kz-\omega t).  \label{h+}
\end{equation}%
One then finds that the time-averaged interaction\ or coupling energy in the
rotating-wave approximation between the electromagnetic and gravitational
radiation fields mediated by the superconductor is%
\begin{equation}
\overline{\Delta E_{int}^{(1)}}=\frac{1}{16}\left( \frac{m_{2}}{m_{2eff}}%
\right) Ne_{2}A_{+}h_{+}r_{0}^{2}.  \label{CouplingEnergy}
\end{equation}%
Note the presence of the factor $N$, which can be very large, since it can
be on the order of Avogadro's number $N_{0}$.

The calculation for the above coupling energy $\overline{\Delta E_{int}^{(1)}%
}$ proceeds along the same lines as that for the Meissner effect of the
superconductor, which is based on the diamagnetism term $H_{dia}$ in the
expansion of the $same$ time-dependent minimal-coupling Hamiltonian, Eq.(\ref%
{TimeDependentHamiltonian}), given by 
\begin{equation}
H_{dia}=\frac{1}{2m_{2eff}}\left\{ e_{2}\mathbf{A}(t)\cdot e_{2}\mathbf{A}%
(t)\right\} .
\end{equation}%
This leads to an energy shift of the system, which, in first-order
perturbation theory, again in the rotating-wave approximation, is given by%
\begin{equation}
\overline{\Delta E_{dia}^{(1)}}=\frac{1}{32m_{2eff}}%
Ne_{2}^{2}A_{+}^{2}r_{0}^{2}.
\end{equation}%
Again, note the presence of the factor $N$, which can be on the order of
Avogadro's number $N_{0}$. \ From this expression, we can obtain the \textit{%
diamagnetic susceptibility} of the superconductor. \ We know from experiment
that the size of this energy shift is sufficiently large to cause a complete
expulsion of the magnetic field from the interior of the superconductor,
i.e., a Meissner effect. \ Hence there must also be a complete reflection of
the electromagnetic wave from the interior of the superconductor, apart from
a thin surface layer of the order of the London penetration depth. \ All
forms of diamagnetism, including the Meissner effect, are purely quantum
effects.

Similarly, there exists a diamagnetic-like ``gravitomagnetic'' term $H_{GM}$%
\ in the expansion of the $same$ minimal-coupling Hamiltonian, Eq.(\ref%
{TimeDependentHamiltonian}), given by%
\begin{equation}
H_{GM}=\frac{1}{2m_{2eff}}\left\{ m_{2}\mathbf{h}(t)\cdot m_{2}\mathbf{h}%
(t)\right\} .
\end{equation}%
This leads to a gravitomagnetic energy shift of the system given in
first-order perturbation theory in the rotating-wave approximation by%
\begin{equation}
\overline{\Delta E_{GM}^{(1)}}=\frac{1}{32m_{2eff}}%
Nm_{2}^{2}h_{+}^{2}r_{0}^{2}.  \label{GMenergy}
\end{equation}%
From this expression, we can obtain the \textit{gravitomagnetic
susceptibility} of the superconductor, which does not vanish. \ This
necessitates the introduction of a nontrivial \textit{constitutive relation}
for the gravitomagnetic field (see Section 6).

\section{The impedance of free space for gravitational plane waves}

It is not enough merely to calculate the coupling energy arising from the
interaction Hamiltonian given by Eq.(\ref{CouplingEnergy}). \ We must also
compare how large this coupling energy is with respect to the free-field
energies of the uncoupled problem, in particular, that of the gravitational
radiation, in order to see how big an effect we expect to see in the
gravitational sector. \ To this end, I shall introduce the concept of 
\textit{impedance matching}, both between the superconductor and free space
in both forms of radiation, and also between the two kinds of waves inside
the superconductor viewed as a transducer. \ The impedance matching problem
determines the\textit{\ efficiency of power transfer} from the antenna to
free space, and from one kind of wave to the other. \ It is therefore useful
to introduce the concept of the \textit{impedance of free space} $Z_{G}$ for
a gravitational plane wave, which is analogous to the concept of the
impedance of free space $Z_{0}$ for an electromagnetic plane wave (here SI
units are more convenient to use than Gaussian cgs units) \cite%
{PanofskyPhillips}%
\begin{equation}
Z_{0}=\frac{E}{H}=\sqrt{\frac{\mu _{0}}{\varepsilon _{0}}}=377\text{ ohms,}
\end{equation}%
where $\mu _{0}$ is the magnetic permeability of free space, and $%
\varepsilon _{0}$ is the dielectric permittivity of free space.

The physical meaning of the ``impedance of free space'' in the
electromagnetic case is that when a plane wave impinges on a large, but
thin, resistive film at normal incidence, due to this film's ohmic losses,
the wave can be absorbed and converted into heat if the resistance per
square element of this film is comparable to 377 ohms \cite{krausAntennas}.
\ In this case, we say that the electromagnetic plane wave has been
approximately ``impedance-matched'' into the film. \ If, however, the
resistance of the thin film is much lower than 377 ohms per square, as is
the case for a superconducting film, then the wave will be reflected by the
film. \ In this case, we say that the wave has been ``shorted out'' by the
superconducting film, and\ that therefore this film reflects electromagnetic
radiation like a mirror. \ By contrast, if the resistance of a normal
metallic film is much larger than 377 ohms per square, then the film is
essentially transparent to the wave. \ As a result, there will be almost
perfect transmission. \ 

The boundary value problem for a travelling plane-wave solution to Maxwell's
equations coupled to a thin resistive film with a resistance per square
element of $Z_{0}/2$, yields a unique solution that\ this is the condition
for the $maximum$ possible fractional absorption of the wave energy by the
film, which is 50\%, along with 25\% of the wave energy being transmitted,
and the remaining 25\% being reflected (see Appendix A) \cite{richards}. \
Under such circumstances, we say that the film has been ``$optimally$
impedance-matched'' to the film. \ This result is valid no matter how thin
the ``thin'' film is.

The gravitomagnetic permeability $\mu _{G}$ of free space is \cite{Landau}%
\cite{Forward}%
\begin{equation}
\mu _{G}=\frac{16\pi G}{c^{2}}=3.73\times 10^{-26}\text{ }\frac{\text{m}}{%
\text{kg}}\text{,}  \label{mu_G}
\end{equation}%
i.e., $\mu _{G}$ is the coupling constant which couples the Lense-Thirring
field to sources of mass current density, in the gravitational analog of
Ampere's law for weak gravity. Ciufolini \textit{et al.} have recently
measured, to within $\pm 20\%$, a value of $\mu _{G}$ which agrees with Eq.(%
\ref{mu_G}), in possibly the first observation of the Earth's Lense-Thirring
field, by means of laser-ranging measurements of the orbits of two
satellites \cite{Ciufolini}. From Eq.(\ref{mu_G}), I find that the impedance
of free space is \cite{kraus}%
\begin{equation}
Z_{G}=\frac{E_{G}}{H_{G}}=\sqrt{\frac{\mu _{G}}{\varepsilon _{G}}}=\mu _{G}c=%
\frac{16\pi G}{c}=1.12\times 10^{-17}\text{ }\frac{\text{m}^{2}}{\text{s}%
\cdot \text{kg}}\text{,}  \label{Z_G}
\end{equation}%
where the fact\ has been used that both electromagnetic and gravitational
plane waves propagate at the same speed 
\begin{equation}
c=\frac{1}{\sqrt{\varepsilon _{G}\mu _{G}}}=\frac{1}{\sqrt{\varepsilon
_{0}\mu _{0}}}=3.00\times 10^{8}\text{ }\frac{\text{m}}{\text{s}}\text{.}
\end{equation}%
Therefore, the gravitoelectric permittivity $\varepsilon _{G}$ of free space
is%
\begin{equation}
\varepsilon _{G}=\frac{1}{16\pi G}=2.98\times 10^{8}\text{ }\frac{\text{kg}%
^{2}}{\text{N}\cdot \text{m}^{2}}\text{.}  \label{epsilon_G}
\end{equation}

Newton's constant $G$ now enters explicitly through the expression for the
impedance of free space $Z_{G}$, into the problem of the interaction of
radiation and matter. \ Note that $Z_{G}$ is an extremely small quantity. \
Nevertheless, it is also important to note that it is not strictly zero. \
Since nondissipative quantum fluids, such as superfluids and
superconductors, can in principle have strictly zero losses, they can behave
like ``short circuits'' for gravitational radiation. \ Thus we expect that
quantum fluids, in contrast to classical fluids, can behave like perfect
mirrors for gravitational radiation. \ That $Z_{G}$ is so small explains why
it is so difficult to couple $classical$ matter to gravity waves. \ It is
therefore natural to consider using nondissipative $quantum$ matter instead
for achieving an efficient coupling. \ 

By analogy with the electromagnetic case, the physical meaning of the
``impedance of free space'' $Z_{G}$ is that when a gravitational plane wave
impinges on a large, but thin, $viscous$ fluid film at normal incidence, due
to this film's dissipative losses, the wave can be absorbed and converted
into heat, if the dissipation per square element of this film is comparable
to $Z_{G}$. \ Again in this case, we say that the gravitational plane wave
has been approximately ``impedance-matched'' into the film. \ If, however,
the dissipation of the thin film is much lower than $Z_{G}$, as is the case
for nondissipative quantum fluids, then the wave will be reflected by the
film. \ In this case, we say that the wave has been ``shorted out'' by the
superconducting or superfluid film, and that therefore the film should
reflect gravitational radiation like a mirror. \ By contrast, if the
dissipation of the film is much larger than $Z_{G}$, as is the case for
classical matter, then the film is essentially transparent to the wave, and
there will be essentially perfect transmission. \ 

The same boundary value problem holds for a travelling plane-wave solution
to the Maxwell-like equations coupled to a thin viscous fluid film with a
dissipation per square element of $Z_{G}/2$, and yields the same unique
solution that\ this is the condition for the $maximum$ possible fractional
absorption (and the consequent conversion into heat) of the wave energy by
the film, which is 50\%, along with 25\% of the wave energy being
transmitted, and the remaining 25\% being reflected (see Appendix A). \
Under such circumstances, we again say that the film has been ``$optimally$
impedance-matched'' to the film. \ Again, this result is valid no matter how
thin the ``thin'' film is.

When the superconductor is viewed as a transducer, the conversion from
electromagnetic to gravitational wave energy, and vice versa, can be viewed
as an $effective$ dissipation mechanism, where instead of being converted
into heat, one form of wave energy is converted into the other form,
whenever impedance matching is achieved within a thin layer inside the
superconductor. \ As we shall see, this can occur naturally when the
electromagnetic wave impedance is $exponentially$ reduced in extreme type II
superconductors as the wave penetrates into the superconductor, so that a
layer is automatically reached in its interior where the electromagnetic
wave impedance is reduced to a level comparable to $Z_{G}$. \ Under such
circumstances, we should expect efficient conversion from one form of wave
energy to the other.

\section{Maxwell-like equations for gravity waves}

For obtaining the impedance of free space $Z_{0}$ for electromagnetic plane
waves, we recall that one starts from Maxwell's equations \cite%
{PanofskyPhillips}%
\begin{equation}
\mathbf{\nabla \cdot D}=+\rho _{e}
\end{equation}%
\begin{equation}
\mathbf{\nabla \times E=-}\frac{\partial \mathbf{B}}{\partial t}
\end{equation}%
\begin{equation}
\mathbf{\nabla \cdot B}=0
\end{equation}%
\begin{equation}
\mathbf{\nabla \times H=}+\mathbf{j}_{e}+\frac{\partial \mathbf{D}}{\partial
t},
\end{equation}%
where $\rho _{e}$ is the electrical free charge density (here, the charge
density of Cooper pairs), and $\mathbf{j}_{e}$ is the electrical current
density (due to Cooper pairs), $\mathbf{D}$ is the displacement field, $%
\mathbf{E}$ is the electric field,\ $\mathbf{B}$ is the magnetic induction
field, and $\mathbf{H}$ is the magnetic field intensity. \ The constitutive
relations (assuming an isotropic medium) are%
\begin{equation}
\mathbf{D}=\kappa _{e}\varepsilon _{0}\mathbf{E}
\end{equation}%
\begin{equation}
\mathbf{B}=\kappa _{m}\mu _{0}\mathbf{H}
\end{equation}%
\begin{equation}
\mathbf{j}_{e}=\sigma _{e}\mathbf{E,}
\end{equation}%
where $\kappa _{e}$ is the dielectric constant of the medium, $\kappa _{m}$
is its relative permeability, and $\sigma _{e}$ is its electrical
conductivity. \ We then convert Maxwell's equations into wave equations for
free space in the usual way, and conclude that the speed of electromagnetic
waves in free space is $c=(\varepsilon _{0}\mu _{0})^{-1/2}$, and that the
impedance of free space is $Z_{0}=(\mu _{0}/\varepsilon _{0})^{1/2}$. \ The
impedance-matching problem of a plane wave impinging on a\ thin, resistive
film is solved by using standard boundary conditions in conjunction with the
constitutive relation $\mathbf{j}_{e}=\sigma _{e}\mathbf{E}$.

Similarly, for $weak$ gravity and $slow$ matter, Maxwell-like equations have
been derived from the linearized form of Einstein's field equations \cite%
{Forward}\cite{Braginsky}\cite{Becker}\cite{Tajmar}. \ The gravitoelectric
field $\mathbf{E}_{G}$, which is identical to the local acceleration due to
gravity $\mathbf{g}$,\ is analogous to the electric field $\mathbf{E}$, and
the gravitomagnetic field $\mathbf{B}_{G}$, which is identical to the
Lense-Thirring field, is analogous to the magnetic field $\mathbf{B}$; they
are related to the vector potential $\mathbf{h}$\ in the radiation gauge as
follows:%
\begin{equation}
\mathbf{g=-}\frac{\partial \mathbf{h}}{\partial t}\text{ and }\mathbf{B}_{G}%
\mathbf{=\nabla \times h}\text{ , }  \label{g}
\end{equation}%
which correspond to the electromagnetic relations in the radiation gauge%
\begin{equation}
\mathbf{E=-}\frac{\partial \mathbf{A}}{\partial t}\text{ and }\mathbf{%
B=\nabla \times A}\text{ .}
\end{equation}%
The physical meaning of $\mathbf{g}$ is that it is the three-acceleration of
a local, freely-falling test particle induced by the gravitational
radiation, as seen by an observer in a local inertial frame located at the
center of mass of the superconductor.\ \ The local three-acceleration $%
\mathbf{g}=-\partial \mathbf{h}/\partial t$ is the local time derivative\ of
the local three-velocity $-\mathbf{h}$ of this test particle, which is a
member of a system of noninteracting, locally freely-falling,\ classical
test particles (e.g., interstellar dust) with a velocity field $\mathbf{v}%
_{test}(x,y,z,t)=-\mathbf{h}(x,y,z,t)$ as viewed by an observer in the
center-of-mass inertial frame. Similarly, the physical meaning of the
gravitomagnetic field $\mathbf{B}_{G}$ is that it is the local angular
velocity of an inertial frame centered on the same test particle, with
respect to the same observer's inertial frame, which is centered on the
freely-falling center-of-mass of the superconductor. Thus $\mathbf{B}_{G}$\
is the Lense-Thirring field induced by gravitational radiation. \ 

The Maxwell-like equations for weak gravitational fields (upon setting the
PPN (``Parametrized Post-Newton'') parameters to be those of general
relativity) are \cite{Braginsky}%
\begin{equation}
\mathbf{\nabla \cdot D}_{G}=-\rho _{G}  \label{Gauss-like}
\end{equation}%
\begin{equation}
\mathbf{\nabla \times g=-}\frac{\partial \mathbf{B}_{G}}{\partial t}
\label{Faraday-like}
\end{equation}%
\begin{equation}
\mathbf{\nabla \cdot B}_{G}=0
\end{equation}%
\begin{equation}
\mathbf{\nabla \times H}_{G}\mathbf{=}-\mathbf{j}_{G}+\frac{\partial \mathbf{%
D}_{G}}{\partial t}  \label{Ampere-like}
\end{equation}%
where $\rho _{G}$ is the density of local rest mass in the local rest frame
of the matter, and $\mathbf{j}_{G}$ is the local rest-mass current density
in this frame (in the case of classical matter, $\mathbf{j}_{G}=\rho _{G}%
\mathbf{v}$, where $\mathbf{v}$ is the coordinate three-velocity of the
local rest mass; in the quantum case, see Eq.(\ref{current density})). \
Here $\mathbf{H}_{G}$ is the gravitomagnetic field intensity, and $\mathbf{D}%
_{G}$ is the gravitodisplacement field.

Again, converting the Maxwell-like equations for weak gravity into a wave
equation for free space in the standard way, we conclude that the speed of
GR waves in free space is $c=(\varepsilon _{G}\mu _{G})^{-1/2}$, which is
identical in GR to the vacuum speed of light, and that the impedance of free
space for GR waves is $Z_{G}=(\mu _{G}/\varepsilon _{G})^{1/2}$, whose
numerical value is given by Eq.(\ref{Z_G}). \ Since the forms of these
equations are identical to those of Maxwell's equations, the same boundary
conditions follow from them, and therefore the same solutions for
electromagnetic problems carry over formally to the gravitational ones. \
These include the solution for\ the optimal impedance-matching problem for a
thin, dissipative film given in Appendix A.

At this point, I would like to introduce the following constitutive
relations (assuming an isotropic medium), which are analogous to those in
Maxwell's theory, viz.,%
\begin{equation}
\mathbf{D}_{G}=4\kappa _{GE}\varepsilon _{G}\mathbf{g}  \label{kappa_GE}
\end{equation}%
\begin{equation}
\mathbf{B}_{G}=\kappa _{GM}\mu _{G}\mathbf{H}_{G}  \label{kappa_GM}
\end{equation}%
\begin{equation}
\mathbf{j}_{G}=-\sigma _{G}\mathbf{g}  \label{sigma_G}
\end{equation}%
where $\varepsilon _{G}$ is the gravitoelectric permittivity of free space
given by Eq.(\ref{epsilon_G}), $\mu _{G}$ is the gravitomagnetic
permeability of free space given by Eq.(\ref{mu_G}), $\kappa _{GE}$ is the
gravitoelectric dielectric constant of a medium, $\kappa _{GM}$ is its
gravitomagnetic relative permeability, and $\sigma _{G}$ is the
gravitational analog of the electrical conductivity of the medium, whose
magnitude is inversely proportional to its viscosity. \ It is natural to
choose to define the constitutive relation, Eq.(\ref{sigma_G}), with a minus
sign, so that for $dissipative$ media, $\sigma _{G}$ is always a $positive$
quantity. \ That this constitutive relation should be introduced here is
motivated by the fact that gravitational radiation produces a \textit{shear
field} in the quadrupolar transverse velocity field of test particles given
by Eq.(\ref{h+}), so that one would expect that the \textit{shear viscosity}
of the medium through which it propagates should enter into its dissipation.
\ Otherwise, there could never be any dissipation of gravitational radiation
as it propagates through a medium. \ The factor of 4 on the right hand side
of Eq.(\ref{kappa_GE}) implies that Newton's law of universal gravitation
emerges from Einstein's theory of GR in the correspondence principle limit.
\ In this limit, the weak equivalence principle is recovered by demanding
that $\kappa _{GE}$ approach unity at low frequencies.

The phenomenological parameters $\kappa _{GE},$ $\kappa _{GM},$ and $\sigma
_{G}$ must be determined by experiment. \ Because of the possibility of
large Meissner-like effects such as in superconductors and other quantum
fluids, $\kappa _{GM}$ can differ substantially from $1$ at low frequencies
(see Appendix B). \ The nonvanishing gravitomagnetic susceptibility
calculated from Eq.(\ref{GMenergy}) should lead to a value of $\kappa _{GM}$
compatible with such large Meissner-like effects. \ Also, note that $\kappa
_{GM}$ can be spatially inhomogeneous, such as near the surface of a
superconductor.

But why is it even $permissible$ to introduce nontrivial constitutive
relations, Eqs.(\ref{kappa_GE}), (\ref{kappa_GM}), and (\ref{sigma_G}), with 
$\kappa _{GE}\neq 1$, $\kappa _{GM}\neq 1$, and $\sigma _{G}\neq 0$? \ One
argument against the introduction of such relations is that the sources of
all gravitational fields, including the gravitoelectric and gravitomagnetic
fields, are entirely determined by the masses and the mass currents of the
material medium in a \textit{composition-independent} way, since the source
of spacetime curvature in Einstein's field equations arise solely from the
stress-energy tensor, which includes the energy density, momentum density,
and stress associated with all forms of matter and all nongravitational
fields, and which is coupled to spacetime solely via $G$. \ Thus the source
of spacetime curvature must be independent of the composition of the
material sources, and of the kind of nongravitational interaction which
binds the material together. \ In this line of reasoning, there cannot be
any mysterious property of the material which would make $\kappa _{GE}$, $%
\kappa _{GM}$, and $\sigma _{G}$\ different for different materials. \ The
only permissible values of these constants would then be $\kappa _{GE}=1$, $%
\kappa _{GM}=1$, and $\sigma _{G}=0$. \ Otherwise, there would be a seeming
violation of the equivalence principle, since there would then be a \textit{%
composition-dependent} response of different kinds of matter to
gravitational fields.

However, one must carefully distinguish between the weak equivalence
principle, which has been extensively experimentally tested in the
low-frequency, \textit{Newtonian-gravity} limit, and what I shall call the
``extended'' equivalence principle, which extends the
composition-independence of the weak equivalence principle to include\ the
response, in particular, the \textit{linear response}, of all kinds of
matter to \textit{Post-Newtonian gravitational fields}, such as to
Lense-Thirring and gravitational radiation fields. \ Because of the
difficulty of generating and detecting these Post-Newtonian fields, this
``extended'' equivalence principle has not been experimentally tested. \ 

One implication of the ``extended'' equivalence principle would be that 
\begin{equation}
\kappa _{GE}=1,\text{ }\kappa _{GM}=1,\text{and }\sigma _{G}=0,
\label{GeneralizedEquivalencePrinciple}
\end{equation}%
in the linear response of all matter to gravitational radiation at all
frequencies. \ Suppose that this ``extended'' equivalence principle were
true. \ I shall argue below using the Kramers-Kronig relations, in
particular, the zero-frequency sum rule which follows from these relations,
that this would lead to a conflict with some known observational facts.

The plane-wave solution of the above Maxwell-like equations lead to the
propagation of a gravitational plane wave at a frequency $\omega $ at a
phase velocity $v_{phase}(\omega )$ given by%
\begin{equation}
v_{phase}(\omega )=c/n_{G}(\omega ),  \label{PhaseVelocity}
\end{equation}%
where the index of refraction $n_{G}(\omega )$ of the medium for the
gravitational plane wave is given by%
\begin{equation}
n_{G}(\omega )=\left( \kappa _{GE}(\omega )\kappa _{GM}(\omega )\right)
^{1/2}.
\end{equation}%
Such a solution is formally identical to that for an electromagnetic plane
wave propagating inside a dispersive optical medium, whose index of
refraction is given by $n(\omega )=\left( \kappa _{e}(\omega )\kappa
_{m}(\omega )\right) ^{1/2}$. \ Since we are considering the\ \textit{linear
response} of the medium to $weak$ gravitational radiation fields, and since
the response of the medium must be $causal$, the index of refraction $%
n_{G}(\omega )$ must obey the Kramers-Kronig relations \cite{Ditchburn}%
\begin{equation}
\mbox{\rm{Re} }n_{G}(\omega )-1=\frac{1}{\pi }P\int_{-\infty }^{+\infty }\frac{%
\mbox{\rm{Im} }n_{G}(\omega ^{\prime })}{\omega ^{\prime }-\omega }d\omega
^{\prime }
\end{equation}%
\begin{equation}
\mbox{\rm{Im} }n_{G}(\omega )=-\frac{1}{\pi }P\int_{-\infty }^{+\infty }\frac{%
\mbox{\rm{Re} }n_{G}(\omega ^{\prime })-1}{\omega ^{\prime }-\omega }d\omega
^{\prime },  \label{K-K2}
\end{equation}%
where $P$ denotes the Cauchy Principal Value. \ The zero-frequency sum rule
follows from the first of these relations, viz.,%
\begin{equation}
\mbox{\rm{Re} }n_{G}(\omega \rightarrow 0)=1+\frac{c}{\pi }\int_{0}^{+\infty }%
\frac{\alpha _{G}(\omega ^{\prime })}{\left( \omega ^{\prime }\right) ^{2}}%
d\omega ^{\prime },  \label{Zero-Frequency-Sum-Rule}
\end{equation}%
where $\alpha _{G}(\omega )$ is the power attenuation coefficient of the
gravitational plane wave at frequency $\omega $ propagating through the
medium, i.e., $\exp (-\alpha _{G}(\omega )z)$ is the exponential factor
which attenuates the power of a wave propagating along the $z$ axis. \ The
nonvanishing dissipation coefficient $\sigma _{G}(\omega )$ introduced in
conjunction with the constitutive relation Eq.(\ref{sigma_G}) will lead to a
nonvanishing value of $\alpha _{G}(\omega )$. \ Assuming that the medium is
in its ground state, the gravity wave cannot grow exponentially with
propagation distance $z$. \ Hence $\alpha _{G}(\omega )>0$ for all
frequencies $\omega $, and therefore%
\begin{equation}
\mbox{\rm{Re} }n_{G}(\omega \rightarrow 0)>1.
\end{equation}

Now suppose that the ``extended'' equivalence principle were true. \ Then at
all frequencies $\kappa _{GE}(\omega )=1$ and $\kappa _{GM}(\omega )=1$,
independent of the composition of the medium. \ In particular at low
frequencies, this implies that the index of refraction%
\begin{equation}
n_{G}(\omega \rightarrow 0)=1  \label{n_G(0)=1}
\end{equation}%
must strictly\ be unity. \ It then follows from the above zero-frequency sum
rule, that the attenuation coefficient%
\begin{equation}
\alpha _{G}(\omega )=0\text{ for all }\omega  \label{alpha=0}
\end{equation}%
must strictly vanish for all frequencies $\omega $. \ This result would
imply that absorption of gravitational radiation would be impossible at any
frequency by any kind of matter. \ Detectors of gravitational radiation
would be impossible due to this ``extended'' equivalence principle. \ By
reciprocity, emission of gravitational radiation by any kind of matter at
any frequency would likewise also be impossible. \ Gravitational radiation
might as well not exist \cite{Loinger}. \ This, however, is contradicted by
the observations of Taylor and Weisberg \cite{Taylor}. \ Thus the
``extended'' equivalence principle must be a false extension of the weak
equivalence principle. \ Hence, not only is it $permissible$, but it is also 
$necessary$, to introduce constitutive equations with nontrivial values of $%
\kappa _{GE}$, $\kappa _{GM}$, and $\sigma _{G}$.

It should be stressed here that although the above Maxwell-like equations
look formally identical to Maxwell's (apart from a sign change of the source
terms), there is an elementary physical difference between gravity and
electricity, which must not be overlooked. \ In electrostatics, the
existence of both signs of charges means that both repulsive and attractive
forces are possible, whereas in gravity, only positive signs of masses, and
only attractive gravitational forces between masses, are observed. \ One
consequence of this experimental fact is that whereas\ it is possible to
construct Faraday\ cages that completely screen out electrical forces, and
hence electromagnetic radiation fields, it is impossible to construct
gravitational analogs of such Faraday cages that screen out gravitoelectric
forces, such as Earth's gravity.

However, the gravitomagnetic force can be either repulsive or attractive in
sign, unlike the gravitoelectric force. \ For example, the gravitomagnetic
force between two parallel current-carrying pipes changes sign, when the
direction of the current flow is reversed in one of the pipes, according to
the Ampere-like law Eq.(\ref{Ampere-like}). \ Hence $both$ signs of this
kind of gravitational force are possible. \ One consequence of this is that
gravitomagnetic forces $can$ cancel out, so that, unlike gravitoelectric
fields, gravitomagnetic fields $can$ in principle be screened out of the
interiors of material bodies. \ A dramatic example of this is the complete
screening out of the Lense-Thirring field by superconductors in a
Meissner-like effect, i.e., the complete expulsion of the gravitomagnetic
field from the interior of these bodies, which is predicted by the
Ginzburg-Landau theory given below in Section 8, and by the Gross-Pitaevskii
theory for atomic BECs given in Appendix B. \ Therefore the expulsion of
gravitational radiation fields by superconductors can also occur, and\ thus
mirrors for this kind of radiation, although counterintuitive, are not
impossible.

\section{Poynting-like vector and the power flow of gravitational radiation}

In analogy with classical electrodynamics, having obtained the impedance of
free space $Z_{G}$, we are now in a position to calculate the time-averaged
power flow in a gravitational plane through a gravitational analog of
Poynting's theorem in the weak-gravity limit. \ The local time-averaged
intensity of a gravitational plane wave is given by the time-averaged
Poynting-like vector%
\begin{equation}
\overline{\mathbf{S_{G}}}=\overline{\mathbf{E}_{G}\mbox{\boldmath $\times$} 
\mathbf{H}_{G}}\text{ .}
\end{equation}%
For a plane wave propagating in the vacuum, the local relationship between
the magnitudes of the $\mathbf{E}_{G}$ and $\mathbf{H}_{G}$ fields is given
by%
\begin{equation}
\left| \mathbf{E}_{G}\right| =Z_{G}\left| \mathbf{H}_{G}\right| \text{ .}
\end{equation}%
From Eq.(\ref{g}a), it follows that the local\ time-averaged intensity,
i.e., the power per unit area, of a harmonic plane wave of angular
frequency\ $\omega $ is given by%
\begin{equation}
\left| \overline{\mathbf{S}_{G}}\right| =\frac{1}{2Z_{G}}\left| \mathbf{E}%
_{G}\right| ^{2}=\frac{\omega ^{2}}{2Z_{G}}\left| \mathbf{h}\right| ^{2}=%
\frac{c^{3}\omega ^{2}}{32\pi G}\left| h_{0i}\right| ^{2}\text{.}
\end{equation}%
For a Gaussian-Laguerre mode of a quadrupolar gravity-wave beam\ propagating
at 10 GHz with an intensity of a milliwatt per square centimeter, the
velocity amplitude $\left| \mathbf{h}\right| $ is typically%
\begin{equation}
\left| \mathbf{h}\right| \simeq 2\times 10^{-20}\text{ m/s, }
\end{equation}%
or the dimensionless strain parameter $\left| h_{0i}\right| =\left| \mathbf{h%
}\right| /c$ is typically%
\begin{equation}
\left| h_{0i}\right| \simeq 8\times 10^{-31}\text{ ,}
\end{equation}%
which is around ten orders of magnitude smaller than the typical strain
amplitudes observable in the earlier versions of LIGO. \ At first sight, it
would seem extremely difficult to detect such tiny amplitudes. \ However, if
the natural impedance matching process in dissipationless, extreme type II
superconductors to be discussed below can be achieved in practice, then both
the generation and the detection of such small strain amplitudes should not
be impossible.

To give an estimate of the size of the magnetic field amplitudes which
correspond to the above gravitational wave amplitudes, one uses energy
conservation in a situation in which the powers in the EM and GR waves
become comparable to each other in the natural impedance-matching process
described below, where, in the special case of perfect power conversion
(i.e., perfect impedance matching) from EM to GR radiation,%
\begin{equation}
\frac{\omega ^{2}}{2Z_{0}}\left| \mathbf{A}\right| ^{2}=\frac{\omega ^{2}}{%
2Z_{G}}\left| \mathbf{h}\right| ^{2}
\end{equation}%
in the free space above the surface of the superconductor, from which it
follows that 
\begin{equation}
\frac{\left| \mathbf{A}\right| }{\left| \mathbf{h}\right| }=\frac{\left| 
\mathbf{B}\right| }{\left| \mathbf{B}_{G}\right| }=\frac{\left| \mathbf{B}%
\right| }{\left| \mathbf{\Omega }_{G}\right| }=\left( \frac{Z_{0}}{Z_{G}}%
\right) ^{1/2}\text{,}
\end{equation}%
where the ratio is given by the square-root of the impedances of free space $%
(Z_{0}/Z_{G})^{1/2}$ instead of\ the mass-to-charge ratio $2m_{2}/e_{2}$
ratio implied by Eq.(\ref{Larmor}). \ Thus for the\ above numbers%
\begin{equation}
\left| \mathbf{B}\right| \simeq 5\times 10^{-3}\text{ Tesla .}
\end{equation}

\section{Ginzburg-Landau equation coupled to both electromagnetic and
gravitational radiation}

A superconductor in the presence of the electromagnetic field $\mathbf{A}(t)$
alone is well described by the Ginzburg-Landau (G-L) equation for the
complex order parameter $\psi $, which in the quantum adiabatic theorem
limit is given by \cite{Tinkham} \ 
\begin{equation}
\frac{1}{2m_{2eff}}\left( \frac{\hbar }{i}\mathbf{\nabla }-e_{2}\mathbf{A}%
(t)\right) ^{2}\psi +\beta |\psi |^{2}\psi =-\alpha \psi .  \label{GL}
\end{equation}%
When $\mathbf{A}$ is time-independent, this equation has the same form as
the time-independent Schr\"{o}dinger equation for a particle (i.e., a Cooper
pair) with mass $m_{2eff}$ and a charge $e_{2}$ with an energy eigenvalue $%
-\alpha $, except that there is an extra nonlinear term whose coefficient is
given by the coefficient $\beta $, which arises at a microscopic level from
the Coulomb interaction between Cooper pairs \cite{Tinkham}. \ The values of
these two phenomenological parameters $\alpha $ and $\beta $ must be
determined by experiment. \ There are two important length scales associated
with the two parameters $\alpha $ and $\beta $ of this equation, which can
be obtained by a dimensional analysis of Eq.(\ref{GL}). \ The first is the 
\textit{coherence length}%
\begin{equation}
\xi =\left( \frac{\hbar ^{2}}{2m_{2eff}|\alpha |}\right) ^{1/2}\text{ ,}
\end{equation}%
which is the length scale on which the condensate charge density $e_{2}|\psi
|^{2}$ vanishes, as one approaches the surface of the superconductor from
its interior, and hence the length scale on which the electric field $%
\mathbf{E}(t)$ is screened inside the superconductor. \ The second is the 
\textit{London penetration depth}%
\begin{equation}
\lambda _{L}=\left( \frac{\hbar ^{2}}{2m_{2eff}\beta |\psi |^{2}}\right)
^{1/2}\rightarrow \left( \frac{\varepsilon _{0}m_{2eff}c^{2}}{e_{2}^{2}|\psi
_{0}|^{2}}\right) ^{1/2}\text{ ,}  \label{LondonDepth}
\end{equation}%
which is the length scale on which an externally applied magnetic field $%
\mathbf{B}(t)=\mathbf{\nabla \times A}(t)$ vanishes due to the Meissner
effect, as one penetrates into the interior of the superconductor away from
its surface. Here\ $|\psi _{0}|^{2}$ is the pair condensate density deep
inside the superconductor, where it approaches a constant.

The G-L equation represents a mean field theory of the superconductor at the
macroscopic level, which can be derived from the underlying microscopic BCS
theory \cite{Gorkov}. \ The meaning of the complex order parameter $\psi
(x,y,z)$ is that it is the Cooper pair condensate wavefunction. \ Since $%
\psi (x,y,z)$ is defined as a complex field defined over \textit{ordinary }$%
(x,y,z)$ \textit{space}, it is difficult to discern at this level of
description the underlying quantum entanglement present in the BCS
wavefunction, which is a many-body wavefunction defined over the \textit{%
configuration space} of the many-electron system.\ \ Nevertheless, quantum
entanglement, and hence instantaneous EPR correlations-at-a-distance, shows
up indirectly\ through the nonlinear term $\beta |\psi |^{2}\psi $, and is
ultimately what is responsible for the Meissner effect (see Appendix B). \
The G-L theory is being used here because it is more convenient than the BCS
theory for calculating the response of the superconductor to
electromagnetic, and also to gravitational, radiation.

\begin{figure}[tbp]
\centerline{\includegraphics{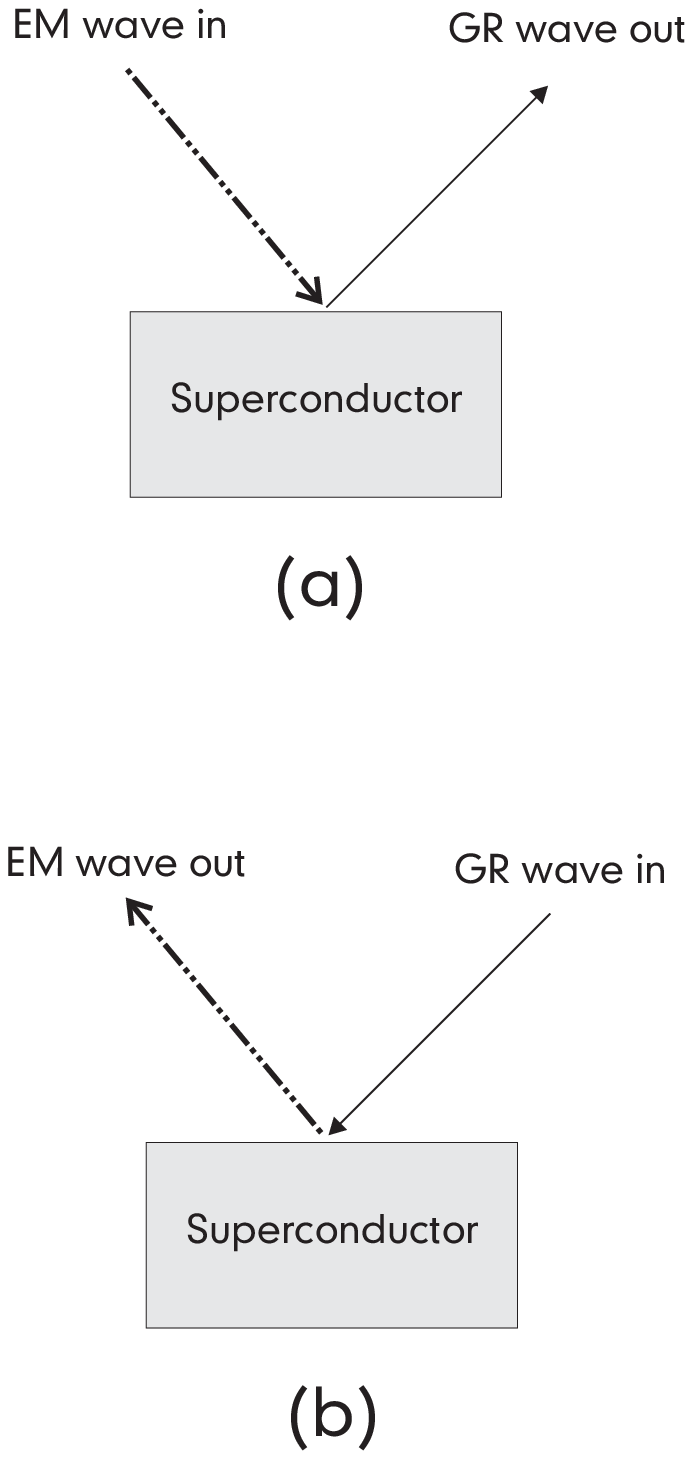}}
\caption{Superconductor as\ an impedance-matched transducer between
electromagnetic (EM) and gravitational (GR) radiation. \ (a) An EM plane
wave is converted upon reflection into a GR plane wave. (b) The reciprocal
(or time-reversed) process in which a GR plane wave is converted upon
reflection into an EM plane wave. \ Both EM and GR waves possess the same 
\textit{quadrupolar} polarization pattern.}
\label{Fresnel}
\end{figure}

I would like to propose that the Ginzburg-Landau equation should be
generalized to include gravitational radiation fields $\mathbf{h}(t)$, whose
frequencies lie well below the BCS gap frequency, by using the
minimal-coupling rule, Eq.(\ref{t-dependent minimal coupling}), to become
the following equation in the quantum adiabatic theorem limit:%
\begin{equation}
\frac{1}{2m_{2eff}}\left( \frac{\hbar }{i}\mathbf{\nabla }-e_{2}\mathbf{A}%
(t)-m_{2}\mathbf{h}(t)\right) ^{2}\psi +\beta |\psi |^{2}\psi =-\alpha \psi .
\label{Generalized GL equation}
\end{equation}%
Again, the ultimate justification for this equation must come from
experiment. \ With this equation, one can predict what happens at the
interface between the vacuum and the superconductor, when both kinds of
radiation are impinging on this surface at an arbitrary angle of incidence
(see Figure \ref{Fresnel}). \ Since there are still only the same two
parameters $\alpha $ and $\beta $ in this equation, there will again be the
same two length scales $\xi $ and $\lambda _{L}$, and \emph{only} these two
length scales, that we had \emph{before} adding the gravitational radiation
term $\mathbf{h}(t)$. \ Since there are no other length scales in this
problem, one would expect that the gravitational radiation fields should
vanish on the same length scales as the electromagnetic radiation fields as
one penetrates deeply into the interior of the superconductor. \ Thus one
expects there to exist a Meissner-like expulsion of the gravitational
radiation fields from the interior of the superconductor.

The meaning of these two length scales becomes clearer when one considers
the extension of the Abrikosov vortex solution of the Ginzburg-Landau
equation with $\mathbf{A}(t)\neq 0$, but with $\mathbf{h}(t)=0$, to the
above generalized Ginzburg-Landau equation with both $\mathbf{A}(t)\neq 0$
and $\mathbf{h}(t)\neq 0$. \ The coherence length $\xi $ is the distance
scale on which the Cooper pair density vanishes near the vortex center, and
the London penetration depth $\lambda _{L}$ is the distance scale on which
the vortical supercurrents decay exponentially away from the center of the
vortex. \ Hence both the gauge fields $\mathbf{A}(t)$ and $\mathbf{h}(t)$
should vanish exponentially on the same scale of length $\lambda _{L}$.

Both $\mathbf{B}(t)$ and $\mathbf{B}_{G}(t)$ fields $must$ vanish into the
interior of the superconductor, since both $\mathbf{A}(t)$ and $\mathbf{h}%
(t) $ fields $must$ vanish in the interior. \ Otherwise, the
single-valuedness of $\psi $ would be violated. \ Suppose that $\mathbf{A}%
(t) $ did not vanish deep inside the superconducting slab, which is
topologically simply connected. \ Then Yang's nonintegrable phase factor for
arbitrary circuits $C $ \cite{Yang1977} 
\begin{equation}
\exp \left( \left( ie_{2}/\hbar \right) \oint_{C}\mathbf{A}(t)\cdot d\mathbf{%
l}\right) ,
\end{equation}%
which is a gauge-independent quantity,\ would also not vanish, which would
lead to a violation of the single-valuedness of $\psi $. \ Suppose that $%
\mathbf{h}(t)$ did not vanish. \ For weak gravity, the nonintegrable phase
factor for small arbitrary circuits $C$ 
\begin{equation}
\exp \left( \left( im_{2}/\hbar \right) \oint_{C}\mathbf{h}(t)\cdot d\mathbf{%
l}\right) ,
\end{equation}%
which is a gauge-independent quantity \cite{MisnerManasse}, would also\ not
vanish, so that again there would be a violation of the single-valuedness of 
$\psi $ (see Figure \ref{circulations}).

The $\mathbf{A}(t)$ and $\mathbf{h}(t)$ fields are coupled strongly to each
other through the $e_{2}\mathbf{A}\cdot \mathbf{h}$ interaction Hamiltonian.
\ Since the electromagnetic interaction is very much stronger than the
gravitational one, the exponential decay of $\mathbf{A}(t)$ on the scale of
the London penetration depth $\lambda _{L}$\ should also govern the
exponential decay of the $\mathbf{h}(t)$ field. \ Thus both $\mathbf{A}(t)$
and $\mathbf{h}(t)$ fields should decay exponentially with the $same$ length
scale $\lambda _{L}$ into the interior of the superconductor. \ This implies
that both EM and GR radiation fields should also be expelled from the
interior, so that a flat surface of this superconductor should behave like a
plane mirror for both EM and GR radiation.\ \ 

The Cooper pair current density $\mathbf{j}$, which acts as the source in
Ampere's law in both the Maxwell and the Maxwell-like equations, can be
obtained in a manner similar to that for the Schr\"{o}dinger equation%
\begin{equation}
\mathbf{j}=\frac{\hbar }{2im_{2eff}}\left( \psi ^{\ast }\mathbf{\nabla }\psi
-\psi \mathbf{\nabla }\psi ^{\ast }\right) -\frac{e_{2}}{m_{2eff}}|\psi |^{2}%
\mathbf{A}-\frac{m_{2}}{m_{2eff}}|\psi |^{2}\mathbf{h}\text{ .}
\label{current density}
\end{equation}%
Note thar $\mathbf{j}$ is nonlinear in $\psi $, but linear in $\mathbf{A}$
and $\mathbf{h}$. \ Near the surface of the superconductor, the gradient
terms dominate, but far into the interior, the $\mathbf{A}$ and the $\mathbf{%
h}$ terms dominate. \ We shall now use $\mathbf{j}$ for calculating the
sources for both Maxwell's equations for the electromagnetic fields, and
also for the Maxwell-like equations for the gravitational fields. \ The
electrical current density, the electrical free charge density, the
rest-mass current density, and the rest mass density, are, respectively,%
\begin{equation}
\mathbf{j}_{e}=e_{2}\mathbf{j}\text{ , }\rho _{e}=e_{2}|\psi |^{2}\text{ , }%
\mathbf{j}_{G}=m_{2}\mathbf{j}\text{ , }\rho _{G}=m_{2}|\psi |^{2}\text{ .}
\end{equation}

I have not yet solved the generalized Ginzburg-Landau equation, Eq.(\ref%
{Generalized GL equation}), coupled to both the Maxwell and Maxwell-like
equations through these currents and densities. \ These coupled equations
are nonlinear in $\psi $, but are linear in $\mathbf{A}$ and $\mathbf{h}$
for weak radiation fields. \ However, from dimensional considerations, I can
make the following remarks. \ The electric field $\mathbf{E}(t)$ should be
screened out exponentially towards the interior of the superconductor on a
length scale set by the coherence length $\xi $, since the charge density $%
\rho _{e}=$ $e_{2}|\psi |^{2}$ vanishes exponentially on this length scale
near the surface of the superconductor. \ Similarly, the magnetic field $%
\mathbf{B}(t)$ should vanish exponentially towards the interior of the
superconductor, but on a different length scale set by the London
penetration depth $\lambda _{L}$. \ Both fields vanish exponentially, but on
different length scales. \ 

At first sight, it would seem that similar considerations would apply to the
gravitational fields $\mathbf{g}(t)$ and $\mathbf{B}_{G}(t)$. \ However,
since there exists only a positive sign of mass for gravity, the
gravitoelectric field $\mathbf{g}(t)$ $cannot$ be screened out, just as in
the case of Earth's gravity. \ Nevertheless, the gravitomagnetic field $%
\mathbf{B}_{G}(t)$ $can$ be, indeed $must$ be, screened by the
quantum-mechanical currents $\mathbf{j}$, in order to preserve the
single-valuedness of $\psi $. \ The quantum-current source terms responsible
for the screening out of the $\mathbf{B}_{G}(t)$ field in Meissner-like
effects are not coupled to spacetime by means of Newton's constant $G$
through the\ right-hand side of the Ampere-like law, Eq.(\ref{Ampere-like}),
but are coupled directly without the mediation of $G$ through the
gravitomagnetic constitutive relation, Eq.(\ref{kappa_GM}), and through the
Faraday-like law, Eq.(\ref{Faraday-like}), whose right-hand side does not
contain $G$. \ For the reasons given above, $\mathbf{B}_{G}(t)$ must decay
exponentially on the same scale of length as the gauge field $\mathbf{A}(t)$
and the magnetic field $\mathbf{B}(t)$, namely the microscopic London
penetration depth $\lambda _{L}$.

This conclusion is reinforced by the fact that a Meissner-like effect is
also predicted to occur in the case of atomic BECs (see Appendix B). From
the Gross-Pitaevskii equation, it follows that the Lense-Thirring field
should also be expelled from the bulk of these $neutral$ quantum fluids,
apart from a thin London penetration depth\ $\lambda _{L}=(8\pi \overline{n}%
a)^{-1/2}$, where $\overline{n}$ is the mean atomic density of the BEC, and $%
a$ is the $S$-wave atom-atom scattering length. \ These atom-atom scattering
events, which lead to the formation of the BEC, entangle the momenta of the
participating atoms, and lead to a London penetration depth $\lambda _{L}$
which is a $microscopic$ length scale, and not an astronomically large, $%
macroscopic$ length scale involving Newton's constant $G$.

The exponential decay into the interior of the superconductor of both EM and
GR waves on the scale of $\lambda _{L}$ means that a flat superconducting
surface should behave like a plane mirror for both electromagnetic and
gravitational radiation. \ However, the behavior of the superconductor as an
efficient $mirror$ is no guarantee that it should also be an efficient $%
transducer$ from one type of radiation to the other. \ For efficient power
conversion, a good transducer impedance-matching process from one kind of
radiation to the other is also required. 

\section{Natural impedance matching in extreme type II superconductors \ }

Impedance matching in a natural transduction process between EM and GR waves
could happen near the surface of extreme type II superconductors, where $\xi
<<\lambda _{L}$, and thus the electric field is screened out much more
quickly on the scale of the coherence length $\xi $, than the magnetic
field, which is screened much more slowly on the scale of the penetration
depth $\lambda _{L}$. \ The EM wave impedance for extreme type II
superconductors should therefore decay on the scale of the coherence length $%
\xi $ much more quickly than the GR wave impedance, which should decay much
more slowly on the scale of the penetration depth $\lambda _{L}$. \ The
high-temperature superconductor YBCO is an example of an extreme type II
superconductor, for which $\xi $ is less than $\lambda _{L}$ by three orders
of magnitude \cite{Zettl}. \ 

The wave impedance $Z=E/H$ of an EM plane wave depends exponentially as a
function of $z$, the distance from the surface into the interior of the
superconductor, as follows:%
\begin{equation}
Z(z)=\frac{E(z)}{H(z)}=Z_{0}\exp (-z/\xi +z/\lambda _{L}).
\end{equation}%
The GR wave impedance $Z_{G}$, however, behaves very differently, because of
the absence of the screening of the gravitoelectric field, so that $E_{G}(z)$
should be constant independent of $z$ near the surface, and therefore%
\begin{equation}
Z_{G}(z)=\frac{E_{G}(z)}{H_{G}(z)}=Z_{G}\exp (+z/\lambda _{L}).
\end{equation}%
Thus the $z$-dependence of the ratio of the two kinds of impedances should
obey the exponential-decay law 
\begin{equation}
\frac{Z(z)}{Z_{G}(z)}=\frac{Z_{0}}{Z_{G}}\exp (-z/\xi ).
\end{equation}%
We must at this point convert the two impedances $Z_{0}$ and $Z_{G}$ to the
same units for comparison. \ To do so, we express $Z_{0}$ in the natural
units of the quantum of surface resistance $R_{0}=h/e^{2}$, where $e$ is the
electron charge. \ Likewise, we express $Z_{G}$ in the corresponding natural
units of the quantum of surface dissipation $R_{G}=h/m^{2}$, where $m$ is
the electron mass. \ Thus we get the $dimensionless$ ratio%
\begin{equation}
\frac{Z(z)/R_{0}}{Z_{G}(z)/R_{G}}=\frac{Z_{0}/R_{0}}{Z_{G}/R_{G}}\exp
(-z/\xi )=\frac{e^{2}/4\pi \varepsilon _{0}}{4Gm^{2}}\exp (-z/\xi ).
\label{zeta}
\end{equation}%
Let us define the ``depth of natural impedance-matching'' $z_{0}$ as the
depth where this dimensionless ratio is unity, and thus natural impedance
matching occurs. \ Then%
\begin{equation}
z_{0}=\xi \ln \left( \frac{e^{2}/4\pi \varepsilon _{0}}{4Gm^{2}}\right)
\approx 97\xi .
\end{equation}%
This result is a robust one, in the sense that the logarithm is very
insensitive to changes in numerical factors of the order of unity in its
argument. \ From this, we conclude that it is necessary penetrate into the
superconductor a distance of\ $z_{0}$, which is around a hundred coherence
lengths $\xi $, for the natural impedance matching process to occur. \ When
this happens, transducer impedance matching occurs automatically, and we
expect that the conversion from electromagnetic to gravitational radiation,
and vice versa, to be an efficient one. \ For example, the London
penetration depth of around 6000 \AA\ in the case of YBCO is very large
compared with $97\xi \simeq $400 \AA\ in this material, so that the
electromagnetic field energy density has not yet decayed by much at this
natural impedance-matching plane $z=z_{0}$, although it is mainly magnetic
in character at this point. \ Therefore the transducer power-conversion
efficiency could be of the order of unity, provided that there is no
appreciable parasitic dissipation of the EM radiation fields in the
superconductor before this point.

The Fresnel-like boundary value problem for plane waves incident on the
surface of the superconductor at arbitrary incidence angles and arbitrary
polarizations (see Figure \ref{Fresnel}) needs to be solved in detail before
these conclusions can be confirmed.

\begin{figure}[tbp]
\centerline{\includegraphics{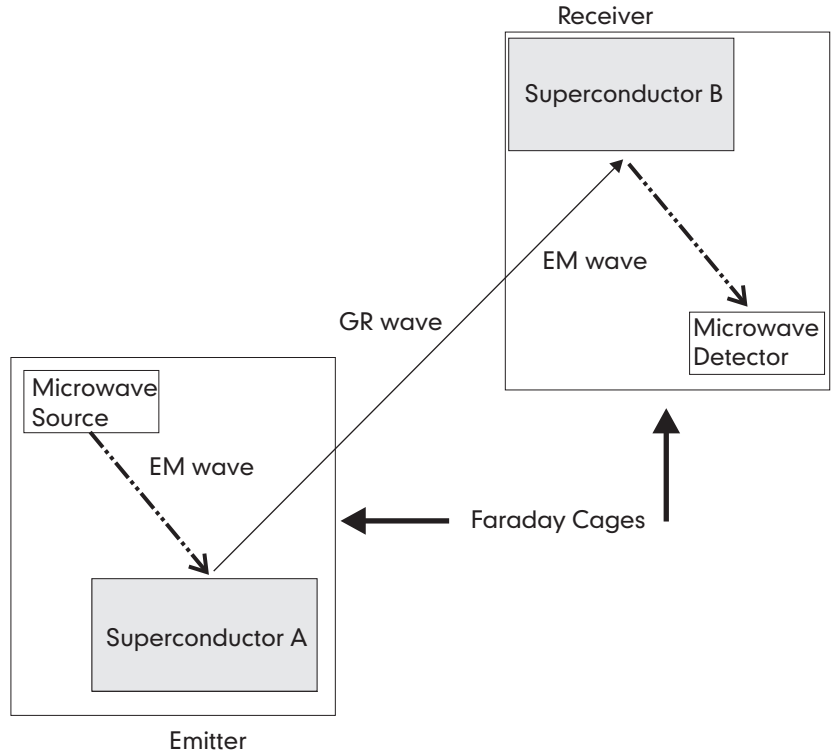}}
\caption{Schematic of a simple Hertz-like experiment, in which gravitational
radiation at 12 GHz could be emitted and received using two superconductors.
\ The ``Microwave Source'' generated, by means of a T-shaped quadrupole
antenna, quadrupolar-polarized electromagnetic radiation at 12 GHz (``EM
wave''), which impinged on Superconductor A (a 1 inch diameter, 1/4 inch
thick piece of YBCO placed inside a dielectric Dewar, i.e., a stack of
styrofoam cups containing liquid nitrogen), and which would be converted
upon reflection into gravitational radiation (``GR wave''). \ The GR wave,
but not the EM wave, could pass through the ``Faraday Cages,'' i.e., normal
metal cans which were lined on the inside with Eccosorb microwave foam
absorbers. \ In the far field of Superconductor A, Superconductor B (also a
1 inch diameter, 1/4 inch thick piece of YBCO in another stack of styrofoam
cups containing liquid nitrogen) would reconvert upon reflection the GR wave
back into an EM wave at 12 GHz, which could then be detected by the
``Microwave Detector,'' which was a sensitive receiver used for microwave
satellite communications, again coupled by another T-shaped quadrupole
antenna to free space. \ The GR wave, and hence the signal at the microwave
detector, should disappear once either superconductor was warmed up above
its transition temperature (90 K), i.e., after the liquid nitrogen boiled
away. }
\label{Hertz}
\end{figure}

\section{Preliminary results of a first experiment, and a proposal for a
future experiment}

However, based on the above crude dimensional and physical arguments, the
prospects for a simple Hertz-like experiment testing these ideas appeared
promising enough that I have performed a first attempt at this experiment
with Walt Fitelson using YBCO at liquid nitrogen temperature. \ The
schematic of this experiment is shown in Figure \ref{Hertz}. \ Details will
be presented elsewhere. \ No observable signal inside the second Faraday
cage was detected, down to a limit of around 70 dB below the microwave power
source of around $-10$ dBm at 12 GHz. \ (We used a commercial satellite
microwave receiver at 12 GHz with a noise figure of 0.6 dB to make these
measurements; the Faraday cages were good enough to block any direct
electromagnetic coupling by more than 70 dB). \ We checked for the presence
of the Meissner effect in the samples\textit{\ in situ} by observing a
levitation effect upon a permanent magnet by these samples at liquid
nitrogen temperature. \ 

Note, however, that since the transition temperature of YBCO is 90 K, there
may have been a substantial ohmic dissipation of the microwaves due to the
remaining normal (unpaired) electrons at our operating temperature of 77 K,
so that the EM wave was absorbed before it could reach the
impedance-matching depth at $z_{0}$. \ It may therefore be necessary to cool
the superconductor down to very low temperatures before the normal electron
component freezes out sufficiently to achieve such an extreme impedance
matching process. \ The exponential decrease of the normal, unpaired
electron population at very low tempertures due to the Boltzmann factor $%
\exp (-E_{g}/k_{B}T)$, where $E_{g}$ is the BCS energy gap at very low
temperatures, and thereby an exponential ``freezing out'' of the ohmic
dissipation of the superconductor, may then allow this impedance matching
process to take place, if no other parasitic dissipative processes remain at
these very low temperatures. \ Assuming that the impedance-matching argument
given in Eq.(\ref{zeta}) is correct, and assuming that in the normal state,
the surface resistance of YBCO is on the order of $h/e^{2}=26$ kilohms per
square in its normal state, one would need a Boltzmann factor of the order
of $e^{-100}$ in order to freeze out the dissipation due to the normal
electrons down to an impedance level comparable to $Z_{G}$. This would imply
that temperatures around a Kelvin should suffice.

However, there exist unexplained residual microwave and far-infrared losses
(of the order of 10$^{-5}$ ohms per square at 10 GHz) in YBCO and other high
T$_{\text{c}}$ superconductors, which are independent of temperature and
have a frequency-squared dependence \cite{Miller}. \ One possible
explanation is that YBCO is a $D$-wave superconductor \cite{Tinkham}, which
is therefore quite unlike the classic, low-temperature $S$-wave
superconductors with respect to their microwave losses. \ In $D$-wave
superconductors, there exists a four-fold symmetry of nodal lines along
which the\ BCS gap vanishes \cite{Davis}. \ Hence it would be impossible to
freeze out the normal, ohmic component of unpaired electrons along these
nodal lines by means of the Boltzmann factor, which makes the choice of YBCO
for the Hertz-like experiment a bad one. \ 

A better choice of superconductor would have been a classic, $S$-wave,
extreme type-II superconductor, such as a niobium alloy which has an
isotropic BCS gap. \ For such a superconductor, the ratio of penetration
depth to coherence length in the ``extreme dirty limit'' is given by De
Gennes \cite{DeGennes}%
\begin{equation}
\kappa \equiv \frac{\lambda _{L}(T)}{\xi (T)}\approx 0.75\left( \frac{%
\lambda _{L}(0)}{\ell }\right) ,
\end{equation}%
where $\ell $ is the mean free path for electron scattering. \ For a niobium
alloy with $\lambda _{L}(0)\approx 400$ \AA\ and $\ell \simeq 1$ \AA , it
follows that $\kappa \simeq 300$. \ Thus impedance matching down to the
extremely low value of $Z_{G}$, as given in Eq.(\ref{Z_G}), would still be
possible. \ However, for achieving such\ an extreme impedance matching
process, one would still need a Boltzmann factor of the order of $e^{-100}$,
and therefore it would be necessary to cool the niobium alloy superconductor
down to very low temperatures, e.g., tens of millikelvins, before the
normal, ohmic electron component would freeze out sufficiently, but more
research needs first to be done to determine whether microwave residual
losses can indeed be exponentially suppressed by such a large exponential
factor. \ An improved Hertz-like experiment using extreme type II
superconductors with extremely low losses, perhaps at millikelvin
temperatures, is a much more difficult, but worthwhile, experiment to
perform.

Such an improved experiment, if successful, would allow us to communicate
through the Earth and its oceans, which, like all classical matter, are
transparent to GR waves. \ Furthermore, it would allow us to directly
observe for the first time the CMB (Cosmic Microwave Background) in GR
radiation, which would tell us much about the very early Universe.

\section{Conclusions}

The conceptual tensions between QM and GR, the two main fields of interest
of John Archibald Wheeler, could indeed lead to important experimental
consequences, much like the conceptual tensions of the past. \ I have
covered here in detail only one of these conceptual tensions, namely, the
tension between the concept of spatial nonseparability of physical systems
due to the notion of nonlocality embedded in the superposition principle, in
particular, in the entangled states of QM, and the concept of spatial
separability of all physical systems due to the notion of locality embedded
in the equivalence principle in GR. \ This has led to the idea of
superconducting antennas and transducers as potentially practical devices,
which could possibly open up a door for further exciting discoveries.

\section{Acknowledgments}

I dedicate this paper to my teacher, John Archibald Wheeler, whose vision
helped inspire this paper. I would like to thank L. Bildsten, S. Braunstein,
C. Caves, A. Charman, M. L. Cohen, P. C. W. Davies, B. S. DeWitt, J. C.
Davis, F. Everitt, W. Fitelson, E. Flanagan, J. C. Garrison, L. Hall, T. H%
\"{a}nsch, J. M. Hickmann, S. Hughes, J. M. Leinaas, R. Laughlin, C.
McCormick, R. Marrus, M. Mitchell, J. Moore, J. Myrheim, R. Packard, R.
Ramos, P. L. Richards, R. Simmonds, D. Solli, A. Speliotopoulos, L.
Stodolsky, G. 't Hooft, C. H. Townes, W. Unruh, S. Weinreb, E. Wright, W.
Wootters, and A. Zettl for helpful discussions. I am grateful to the John
Templeton Foundation for the invitation to contribute to this Volume, and
would like to thank my father-in-law, the late Yi-Fan Chiao, for his
financial and moral support of this work. \ This work was supported also by
the ONR.

\section{Appendix A: Optimal impedance matching of a gravitational plane
wave into a thin, dissipative film}

Let a gravitational plane wave given by Eq.(\ref{h+}) be normally incident
onto a thin, dissipative (i.e., viscous) fluid film. \ Let the thickness $d$
of this film be arbitrarily thin compared to the gravitational analog of the
skin depth $(2/\kappa _{GM}\mu _{G}\sigma _{G}\omega )^{1/2}$, and to the
wavelength $\lambda $. \ The incident fields calculated using Eqs.(\ref{g})\
(here the notation $\mathbf{E}_{G}$ will be used instead of $\mathbf{g}$ for
the gravitoelectric field) are%
\begin{equation}
\mathbf{E}_{G}^{(i)}=-\frac{1}{2}(x,-y,0)\omega h_{+}\sin (kz-\omega t)
\end{equation}%
\begin{equation}
\mathbf{H}_{G}^{(i)}=-\frac{1}{2Z_{G}}(y,x,0)\omega h_{+}\sin (kz-\omega t).
\end{equation}%
Let $\rho $ be the amplitude reflection coefficient for the gravitoelectric
field; the reflected fields from the film are then%
\begin{equation}
\mathbf{E}_{G}^{(r)}=-\rho \frac{1}{2}(x,-y,0)\omega h_{+}\sin (kz-\omega t)
\end{equation}%
\begin{equation}
\mathbf{H}_{G}^{(r)}=+\rho \frac{1}{2Z_{G}}(y,x,0)\omega h_{+}\sin
(kz-\omega t).
\end{equation}%
Similarly the transmitted fields on the far side of the film are%
\begin{equation}
\mathbf{E}_{G}^{(t)}=-\tau \frac{1}{2}(x,-y,0)\omega h_{+}\sin (kz-\omega t)
\end{equation}%
\begin{equation}
\mathbf{H}_{G}^{(t)}=-\tau \frac{1}{2Z_{G}}(y,x,0)\omega h_{+}\sin
(kz-\omega t),
\end{equation}%
where $\tau $ is the amplitude tranmission coefficient. \ The Faraday-like
law, Eq.(\ref{Faraday-like}), and the Ampere-like law, Eq.(\ref{Ampere-like}%
), when applied to\ the tangential components of the gravitoelectric and
gravitomagnetic fields parallel to two appropriately chosen infinitesimal
rectangular loops which straddle the thin film, lead to the two boundary
conditions%
\begin{equation}
\mathbf{E}_{G}^{(i)}+\mathbf{E}_{G}^{(r)}=\mathbf{E}_{G}^{(t)}\text{ and}
\end{equation}%
\begin{equation}
\mathbf{H}_{G}^{(i)}+\mathbf{H}_{G}^{(r)}=\mathbf{H}_{G}^{(t)},
\end{equation}%
which yield the two algebraic relations%
\begin{equation}
1+\rho -\tau =0
\end{equation}%
\begin{equation}
1-\rho -\tau =\left( Z_{G}\sigma _{G}d\right) \tau \equiv \zeta \tau
\end{equation}%
where we have used the constitutive relation, Eq.(\ref{sigma_G}),%
\begin{equation*}
\mathbf{j}_{G}=-\sigma _{G}\mathbf{E}_{G}
\end{equation*}%
to determine the current enclosed by the infinitesimal rectangular loop in
the case of the Ampere-like law, and where we have defined the positive,
dimensionless quantity $\zeta \equiv Z_{G}\sigma _{G}d$, i.e., $Z_{G}$
normalized to the surface dissipation $(\sigma _{G}d)^{-1}$, which is
analogous to the surface resistance, or ``ohms per square'' $(\sigma
_{e}d)^{-1}$ \cite{krausAntennas}. \ The solutions are%
\begin{equation}
\tau =\frac{2}{\zeta +2}\text{ and }\rho =-\frac{\zeta }{\zeta +2}.
\end{equation}%
Using the conservation of energy, we can calculate that the absorptivity $A$%
, i.e., the fraction\ of power absorbed from the incident gravitational wave
and converted into heat, is%
\begin{equation}
A=1-\left| \tau \right| ^{2}-\left| \rho \right| ^{2}=\frac{4\zeta }{(\zeta
+2)^{2}}.
\end{equation}%
To find the condition for maximum absorption, we calculate the derivative $%
dA/d\zeta $ and set it equal to zero. \ The unique solution for maximum
absorptivity occurs at $\zeta =2$, where%
\begin{equation}
A=\frac{1}{2}\text{ and }\left| \tau \right| ^{2}=\frac{1}{4}\text{ and }%
\left| \rho \right| ^{2}=\frac{1}{4}\text{.}
\end{equation}%
Thus the optimal impedance-matching condition into the thin, dissipative
film, i.e., when there exists the maximum rate of conversion of
gravitational wave energy into heat, occurs when the dissipation in the
fluid film is $Z_{G}/2$ per square. \ At this optimum condition, 50\% of the
gravitational wave energy will be converted into heat, 25\% will be
transmitted, and 25\% will be reflected. \ This is true independent of the
thickness $d$ of the film, when the film is very thin. \ This solution is
formally identical to that of the optimal impedance-matching problem of an
electromagnetic plane wave into a thin ohmic film \cite{richards}.

\section{Appendix B: The Gross-Pitaevskii equation and weak gravity}

\subsection{The London penetration depth $\protect\lambda _{L}$\ and the 
\textit{S}-wave scattering length $a$}

The mean-field description of the recently observed atomic BECs is also
given by Eq.(\ref{GL}), except with $\mathbf{A}(t)=0$, whereupon it is
called the ``Gross-Pitaevskii'' (G-P) equation instead of the
``Ginzburg-Landau'' (G-L) equation. \ The G-P equation was first suggested
in the context of a phenomenological theory of superfluid helium, but has
also been applied with great success to atomic BECs \cite{Wiemann}. \ In the
atomic BEC case, it can be derived from a microscopic theory\ of the
weakly-interacting Bose gas due to Bogoliubov. \ (Similarly, the
weakly-interacting photon gas leads to a photonic BEC, and thus a ``photon
superfluid'' \cite{ChiaoPhotonFluid}.) \ Again, the same $two$
phenomenological parameters $\alpha $ and $\beta $ which appear in the
time-independent G-L equation also appear in the time-independent G-P
equation, and therefore also lead to $two$ length scales $\xi $ and $\lambda
_{L}$. \ For the case of the weakly-interacting Bose gases in atomic BECs,
the parameter $\beta $ is directly proportional to the scattering length $a$
for the $S$-wave atom-atom scattering in the BEC at low energies. \ Each
microscopic atom-atom scattering event entangles the momenta of the two
participating atoms, thus producing an entangled state consisting of a sum
of product states of two atoms with opposite momenta. \ This in turn leads
to instantaneous EPR correlations-at-a-distance which violate Bell's
inequality.\ \ Similarly, the Meissner effect, which arises from the $\beta $
term in the G-L equation, also originates microscopically from the quantum
entanglement of pairs of particles participating in individual scattering
events inside the superconductor, which in turn also leads to instantaneous
EPR correlations-at-a-distance.\ 

Using the minimal-coupling rule for neutral particles of mass $m$, which is
a special case of Eq.(\ref{t-dependent minimal coupling}),%
\begin{equation}
\mathbf{p\rightarrow p}-m\mathbf{h}(t),
\end{equation}%
the G-P equation should be generalized in the presence of gravitational
radiation fields to become, in the quantum adiabatic theorem limit, the
equation%
\begin{equation}
\frac{1}{2m_{eff}}\left( \frac{\hbar }{i}\mathbf{\nabla }-m\mathbf{h}%
(t)\right) ^{2}\psi +\beta |\psi |^{2}\psi =-\alpha \psi ,
\label{Generalized Gross-Pitaevskii equation}
\end{equation}%
where $m$ is the vacuum rest mass of the atom, and $m_{eff}$ is its
effective mass. \ In the case of the atomic BECs, the effective mass is
equal to the vacuum rest mass of the atom, i.e., $m_{eff}=m$, to a very good
approximation. \ The minimal-coupling term with the vector potential $%
\mathbf{A}(t)$ is of course absent for these $neutral$ quantum fluids. \
However, as in the G-L equation, the $same$ two parameters $\alpha $ and $%
\beta $ still make their appearance here in the G-P equation$.$ \ The same
dimensional arguments as in Section 8 apply once again, so that, just as in
the generalized G-L equation Eq.(\ref{Generalized GL equation}), two, and 
\textit{only} two, length scales, $\xi $ and $\lambda _{L}$, make their
appearance in the generalized G-P equation, namely, the coherence length 
\begin{equation}
\xi =\left( \frac{\hbar ^{2}}{2m|\alpha |}\right) ^{1/2}=\left( \frac{\hbar
^{2}}{2m|\mu |}\right) ^{1/2}=\left( \frac{1}{8\pi a\overline{n}}\right)
^{1/2}
\end{equation}%
where $\mu =4\pi a\overline{n}\hbar ^{2}/m$ is the chemical potential, $a$
is the scattering length, and $\overline{n}$ is the mean atomic density of
the BEC, and the London penetration depth%
\begin{equation}
\lambda _{L}=\left( \frac{\hbar ^{2}}{2m\beta |\psi |^{2}}\right) ^{1/2},
\end{equation}%
which is derived from the parameter $\beta $; $\lambda _{L}$ is to be
interpreted as the \textit{London penetration depth }of the gravitational
radiation field $\mathbf{h}(t)$, viewed as a gauge-like field,\ into the
interior of the $neutral$ quantum fluid. \ For an atomic BEC, which is a
weakly-interacting Bose gas, the parameter $\beta $ is directly related to
the scattering length $a$ through \cite{Wiemann}%
\begin{equation}
\beta =\frac{4\pi \hbar ^{2}a}{m}.
\end{equation}%
Numerically, for the case of the atomic BEC formed from the rubidium isotope 
$^{87}$Rb, where the scattering length has been measured to be $a=5.77$ nm,
and where the average atomic density is typically $\overline{n}=|\psi
_{0}|^{2}\simeq 10^{14}$ atoms/cm$^{3}$, the London penetration depth is of
the order%
\begin{equation}
\lambda _{L}\approx \left( \frac{\hbar ^{2}}{2m\beta |\psi _{0}|^{2}}\right)
^{1/2}\approx \left( \frac{1}{8\pi a\overline{n}}\right) ^{1/2}\simeq
3\times 10^{-5}\text{ cm,}  \label{BEC-penetration-depth}
\end{equation}%
i.e., about 0.3 $\mu $m, which is a microscopically small length scale. \
For a pure atomic BEC, the coherence length $\xi $ is equal to the London
penetration depth $\lambda _{L}$. \ In general, however, the coherence
length is not equal to the London penetration depth. \ In analogy with
DeGennes's extreme type II superconductivity in alloys for which $\xi
<<\lambda _{L}$, we expect that there exist circumstances under which an
atomic BEC with a cold, dense noble buffer gas would exhibit ``extreme type
II'' behavior, in which the coherence length $\xi $\ is much less than the
London penetration depth $\lambda _{L}$. \ Under these circumstances, it
becomes clear that one should interpret $\lambda _{L}$, and not $\xi $, as
the screening length for quantum currents, and hence the penetration depth
of the gauge-like field $\mathbf{h}(t)$. \ 

This conclusion is reinforced by the fact that the Abrikosov vortex solution
for type II superconductors can be generalized to an Abrikosov-like vortex
solution of the generalized G-P equation Eq.(\ref{Generalized
Gross-Pitaevskii equation}), since the formal structures of these two
problems are the same. \ The meaning of $\lambda _{L}$ in the case of the
Abrikosov-like vortex is that it is the screening distance for $neutral$
quantum currents to exponentially decay from the center of the vortex, and
hence for the \textit{flux quantum} of the gauge-like field $\mathbf{h}$ to
become trapped inside the vortex core, and thus to become well defined.

It is indeed surprising at first sight that Newton's constant $G$ does not
appear here into this expression for the London penetration depth $\lambda
_{L}$, which would have made $\lambda _{L}$ astronomically large. \ For if
the Lense-Thirring field were to arise solely from the mass-current source
term on the right-hand side of the Ampere-like law, Eq.(\ref{Ampere-like}), $%
and$ if the gravitomagnetic relative permeability $\kappa _{GM}$ were simply
that of the vacuum deep inside the BEC, so that $\kappa _{GM}=1$ in the
constitutive relation Eq.(\ref{kappa_GM}) throughout the bulk of the quantum
fluid, then using the substitution $e^{2}/4\pi \varepsilon _{0}\rightarrow
4Gm$ in Eq.(\ref{LondonDepth}), one might expect that $G$ would enter the
London penetration depth in the case of the atomic BEC as follows:%
\begin{equation}
\lambda _{L}^{(G)}=\left( \frac{c^{2}}{16\pi Gm\overline{n}}\right)
^{1/2}\simeq 1.4\times 10^{17}\text{ cm,}
\end{equation}%
i.e., about a tenth of a light-year, which an astronomically large length
scale. \ This would imply that in addition to the two length scales $\xi $
and $\lambda _{L}$ given above, there exists a $third$, and much larger
length scale $\lambda _{L}^{(G)}$, which would somehow also enter into the
solution of the Abrikosov-like vortex problem for extreme type II atomic
BECs, for instance. \ If such were the case, one might expect Meissner-like
expulsions of the Lense-Thirring field on the scale of $\lambda _{L}^{(G)}$
in astrophysical settings, such as in neutron stars \cite{Lano}. \ However,
due to the sign change of the source term of the Ampere-like law, no such
astrophysical Meissner-like effect would in fact exist at all (see the next
section) \cite{Modanese}.

One possible solution to this problem of the inconsistency of length scales
is to introduce a relative permeability $\kappa _{GM}$ which becomes very
large and negative inside the bulk of this neutral quantum fluid, which
implies the existence of a strong Meissner-like effect in which the
Lense-Thirring field is expelled from the interior of the quantum fluid,
apart from a thin layer of thickness $\lambda _{L}$. \ This is a
manifestation of the fact that London's ``rigidity of the macroscopic
wavefunction'' associated with the instantaneous EPR
correlations-at-a-distance originating from the quantum entanglement
characterized by the microscopic scattering length $a$, and hence by the
parameter $\beta $, overpowers the gravitational radiation field, and
prevents its penetration into the quantum fluid (remember that the
gravitational interaction is the weakest of all the interactions). \ When
the cubic nonlinearity $|\psi |^{2}\psi $ associated with the parameter $%
\beta $ in the G-P and G-L equations becomes dominant is also when the
irrotational condition $\mathbf{\nabla \times v}_{\sup }=0$, where $\mathbf{v%
}_{\sup }$ is the superfluid velocity field, becomes valid \cite{putterman}.
\ Hence gravity waves should be reflected from a flat surface of a quantum
fluid, such as superfluid helium or an atomic BEC, as if it were a plane
mirror. \ For otherwise, if the gauge-like $\mathbf{h}(t)$ field with a
weak, but arbitrary, amplitude were allowed to penetrate deeply into the
interior of the quantum fluid, again this would lead to a violation of the 
\textit{single-valuedness }of the complex order parameter or condensate
wavefunction $\psi $ \cite{Abrikosov} (i.e., again consider what would
happen in Figure \ref{circulations} in the case of a neutral atomic BEC,
when gravitational radiation is allowed to penetrate deeply into it with a
wavelength $\lambda <<\lambda _{L}^{(G)}$). \ Thus it is the \textit{%
nonlocal quantum interference} throughout the bulk of the quantum fluid, and
the direct \textit{quantum back-action} upon spacetime of the interference
currents which are produced in response to externally applied Lense-Thirring
fields, which prevents the entry of the Lense-Thirring field, and hence of
gravitational radiation into the bulk of this fluid.

\subsection{The sign and the size of the gravitomagnetic susceptibility $%
\protect\chi _{GM}$ for an atomic BEC}

\bigskip In connection with the strong Meissner-like effect which is
predicted to occur in quantum fluids, it is useful to introduce the concept
of the gravitomagnetic susceptibility $\chi _{GM}$, which is analogous to
the magnetic susceptibility $\chi _{m}$ of electromagnetism, through the
constitutive relation%
\begin{equation}
\mathbf{B}_{G}\equiv \mu _{G}^{\prime }\mathbf{H}_{G}=\kappa _{GM}\mu _{G}%
\mathbf{H}_{G}=\left( 1+\chi _{GM}\right) \mu _{G}\mathbf{H}_{G}.
\label{FullConstitutiveRelation}
\end{equation}%
Thus $\kappa _{GM}=1+\chi _{GM}$, and the gravitomagnetic permeability $\mu
_{G}^{\prime }$\ of a quantum fluid,\ such as an atomic BEC, is given by%
\begin{equation}
\mu _{G}^{\prime }=\left( 1+\chi _{GM}\right) \mu _{G},  \label{mu}
\end{equation}%
which can differ, in principle, from the gravitomagnetic permeability of
free space $\mu _{G}=16\pi G/c^{2}.$ \ One must be careful when considering
the sign of $\chi _{GM}$ that the Ampere-like law, Eq.(\ref{Ampere-like}),\
possesses an opposite sign in front of its mass-current source term relative
to that of the electrical current source term in Ampere's law. \ This sign
change is necessitated by the continuity equation that results from taking
the divergence of this Ampere-like law, in conjunction with the fact that
there must also be a sign change in front of the mass source term of the
Gauss-like law, Eq.(\ref{Gauss-like}), relative to that of electricity: $%
Like $ masses $attract$ each other in ordinary gravity. \ The negative sign
of the mass-current source term implies that two $parallel$ mass currents $%
repel $ each other due to the Lense-Thirring force between two
current-carrying pipes, and that \textit{anti-parallel} mass currents $%
attract$.

Deep inside the bulk of the quantum fluid (e.g., an atomic BEC), the
London-like equation relating the quantum probability current density to the
gauge-like field \textbf{h}, which follows from Eq.(\ref{current density}),
is given in the radiation gauge by%
\begin{equation}
\mathbf{j}=-|\psi _{0}|^{2}\mathbf{h=-}\overline{n}\mathbf{h,}
\end{equation}%
where for atomic BECs, we have used the fact that $m_{eff}=m$. \ This is
analogous to\ the London equation for superconductors, which London used to
derive the Meissner effect, i.e.,%
\begin{equation}
\mathbf{j}=-\frac{e_{2}}{m_{2eff}}|\psi _{0}|^{2}\mathbf{A=-}\frac{e_{2}}{%
m_{2eff}}\overline{n}\mathbf{A.}
\end{equation}%
Let us\ now follow London's procedure, except that the above constitutive
relation Eq.(\ref{FullConstitutiveRelation}) is assumed to hold inside the
medium, so that the magnetostatic Ampere-like law now becomes (using $%
\mathbf{j}_{G}=m\mathbf{j)}$ 
\begin{equation}
\mathbf{\nabla }\times \mathbf{B}_{G}=-\mu _{G}^{\prime }\mathbf{j}_{G}\text{%
,}  \label{StaticAmpere-like}
\end{equation}%
where $\mu _{G}^{\prime }=\left( 1+\chi _{GM}\right) \mu _{G}$, and $\chi
_{GM}$ may be large and negative.\ \ It follows in the radiation gauge, that
one obtains a Yukawa-like equation%
\begin{equation}
\mathbf{\nabla }\times \left( \mathbf{\nabla }\times \mathbf{h}\right) =%
\mathbf{\nabla }\left( \mathbf{\nabla \cdot h}\right) -\nabla ^{2}\mathbf{h}=%
\mathbf{-}\nabla ^{2}\mathbf{\mathbf{h=-}}\frac{1}{\lambda _{L}^{2}}\mathbf{%
\mathbf{h}}\text{ ,}
\end{equation}%
where $\lambda _{L}=\left( -\mu _{G}^{\prime }\overline{n}m\right) ^{-1/2}$
is the London penetration depth. \ There results an exponential decay of the 
$\mathbf{h}$ field starting from the surface towards the interior of the
quantum fluid, i.e., a Meissner-like effect, on the scale of $\lambda _{L}$.
\ Since the London penetration depth\ $\lambda _{L}$ is fixed by \textit{S}%
-wave scattering to be $\left( 8\pi \overline{n}a\right) ^{-1/2}$ where $a$
is the $microscopic$ atomic scattering length, it follows that the
gravitomagnetic susceptibility of the atomic BEC is very large and negative,
so that to a good approximation, $1+\chi _{GM}\rightarrow \chi _{GM}$, and\
thus%
\begin{equation}
\chi _{GM}\approx -\frac{c^{2}}{2G}\frac{a}{m}\simeq -3\times 10^{42}.
\end{equation}%
One consequence of this surprising result is that in Post-Newtonian gravity,
there should exist a very large quantum source term for the Lense-Thirring
field which originates not from the usual classical mass currents coupled to
spacetime through $G$, but from nonlocal quantum interference currents
coupled directly to spacetime without the mediation of $G$. \ There seems to
be no experimental evidence against the existence of such a quantum source
term, and, indeed, the Hess-Fairbank effect (see below) may be evidence for
its existence. \ The implications of the Kramers-Kronig relations in light
of this result would need further exploration.

By contrast, if one were to insist on setting the gravitomagnetic
susceptibility $\chi _{GM}$ equal to zero, so that $\mu _{G}^{\prime }=\mu
_{G}$, i.e., that such a quantum source term does not exist, then due to the
sign change in the source term of the Ampere-like equation Eq.(\ref%
{StaticAmpere-like}), one would obtain a \textit{Helmholtz-like} equation%
\begin{equation}
\nabla ^{2}\mathbf{h}+\left( \frac{1}{\lambda _{L}^{(G)}}\right) ^{2}\mathbf{%
\mathbf{h}}=0,  \label{Helmholtz-like}
\end{equation}%
which has the form $\left( \nabla ^{2}+k^{2}\right) \mathbf{\mathbf{h}}=0$,
instead of the \textit{Yukawa-like} equation%
\begin{equation}
\nabla ^{2}\mathbf{\mathbf{h-}}\left( \frac{1}{\lambda _{L}}\right) ^{2}%
\mathbf{\mathbf{h}}=0,  \label{Yukawa-like}
\end{equation}%
which has the form $\left( \nabla ^{2}-\kappa ^{2}\right) \mathbf{\mathbf{h}}%
=0$. \ The Helmholtz-like equation Eq.(\ref{Helmholtz-like}) possesses
sinusoidal, propagative solutions, rather than exponential, evanescent
solutions of the Yukawa-like equation Eq.(\ref{Yukawa-like}), so that no
Meissner-like effect would exist at all, even in astrophysical settings. \
Therefore any quantum fluid, including superfluid helium and atomic BECs,
would never satisfy, under any circumstances involving gravitational
radiation, the irrotational condition $\mathbf{\nabla \times v}_{\sup }=0$ %
\cite{putterman}: This would lead to a violation of the single-valuedness of 
$\psi $, in contradiction with QM.

\subsection{The Hess-Fairbank experiment \ }

There may already be an experimental hint of the existence of a strong
Meissner-like effect in the case of superfluid helium. \ The Hess-Fairbank
experiment \cite{Hess-Fairbank} is an analog of the Meissner effect seen in
a slowly rotating bucket of liquid helium as it is slowly cooled through the
lambda transition temperature, in which the angular momentum of the fluid is 
$expelled$ from the superfluid below this transition. \ This may be viewed
as evidence for a true Meissner-like effect, in which the Lense-Thirring
field from distant matter in the Universe is expelled. \ From an application
of the limited form of Mach's principle in GR using the Lense-Thirring
effect from distant, rotating shells of stars \cite{MTW}, it follows that
the parabolic meniscus of any steadily rotating classical fluid, including
that of normal liquid helium above the transition temperature, is actually a
consequence of the presence of the Lense-Thirring field from this distant
matter. \ The Hess-Fairbank effect could then be reinterpreted as a true
Meissner-like effect, in which the Lense-Thirring field of these distant
stars is actually expelled. \ Moreover, one could interpret this effect as
evidence for a violation of the ``extended'' equivalence principle discussed
in Section 6, since the response of a quantum fluid (viz., superfluid
helium) to the Lense-Thirring field of distant matter is obviously different
from the\ composition-independent, parabolic-meniscus response of all
classical fluids. \ 

One implication is that the parabolic meniscus should become a flat meniscus
below this thermodynamic transition, except for a thin layer of thickness $%
\lambda _{L}$, which, in general, differs from $\xi $. \ It should be
possible to measure the curvature of the surface of the rotating fluid by
reflecting a laser beam from it. \ The flattening on the meniscus of
rotating liquid helium below the lambda point of a slowly rotating bucket of
this liquid, as it is slowly cooled through the lambda transition, can be
observed by means of laser interferometry \cite{FairbanksLaser}. The flat
meniscus also predicted by Eq.(\ref{Generalized Gross-Pitaevskii equation})
for the slowly rotating bucket of superfluid helium below the transition
temperature implies that this quantum fluid below its transition temperature
should become a \textit{plane mirror} for gravitational radiation. \
However, there is no way of knowing this without a source and detector of
such radiation. \ Hence the experiment on superconductors as transducers
must be performed first. \


\begin{thebibliography}{99}
\bibitem{CurrentCorrelation} Whereas the $longitudinal$ density-density
correlation function is given by a Green function retarded by the speed of
sound, the $transverse$ current-current correlation function, which is
related to vortex motions, is not so retarded, due the irrotational
condition $\mathbf{\nabla \times v}_{\sup }=0$ (see next footnote).
\bibitem{Abrikosov} It should be emphasized that the single-valuedness of
the wavefunction is a principle of quantum mechanics which holds only in the
local inertial frame centered on the center of mass of the system (e.g., the
superconductor) being considered. \ Before turning on the gravitational
radiation field depicted in Figure \ref{circulations}, the initial
conditions are that the superconductor is in its ground state, and in a
state of free fall. \ After the gravitational radiation field is turned on,
the amplitude of this field is assumed to be weak but otherwise arbitrary,
so that no vortices (e.g., Abrikosov vortices in the case of type II
superconductors) are allowed to enter into the superconductor.
\bibitem{homogeneous} We shall later include the effects of spatial
inhomogeneity near the surface of the superconductor by means of the
Ginzburg-Landau equation, Eq.(\ref{Generalized GL equation}).
\bibitem{richards} I thank my colleague, Paul Richards, for pointing to me
out the solution of the optimal-impedance matching problem for
electromagnetic plane waves, whose extension to gravitational plane waves is
presented in Appendix A.
\bibitem{Free-Bound-Sources} In analogy with Maxwell's theory, where we must
distinguish between free and bound charges and charge currents are sources,
we in general must distinguish here also between free and bound masses and
mass currents sources (``free'' here means ``freely falling''), and it is
understood that in the Gauss-like and Ampere-like equations, Eqs.(\ref%
{Gauss-like}) and (\ref{Ampere-like}), the sources refer to $free$ masses
and $free$ mass currents.
\bibitem{Zettl} For YBCO (Yttrium Barium Copper Oxide) superconductors, $\xi 
$ is around 3 to 6 \AA\ along the c-axis, and $\lambda _{L}$ along the
c-axis is around 4000 to 8000 \AA . \ I thank my colleague, Alex Zettl, for
providing me with a useful table of the measured superconducting parameters
for YBCO.
\end{thebibliography}
\end{document}